\documentclass[aps,jcp,amsmath,amssymb,reprint]{revtex4-1}

\usepackage{chemformula} % Formula subscripts using \ch{}
\usepackage[T1]{fontenc} % Use modern font encodings
\usepackage{multirow}
\usepackage{dcolumn}
\usepackage{tabularx}
\usepackage{float}
\usepackage{xr}
\usepackage{hyperref}
\usepackage[utf8]{inputenc}
\usepackage{mciteplus}
\usepackage{subfig}
\usepackage[utf8]{inputenc}
\setcitestyle{super}

\newcommand{\comment}[1]{}

\begin{document}
	\preprint{AIP/123-QED}
	
	\title{Assessing the Tamm-Dancoff Approximation, singlet-singlet, and singlet-triplet excitations with the latest long-range corrected double-hybrid density functionals}
	
	\author{Marcos Casanova-P\'aez}
	\author{Lars Goerigk}%
	\email{lars.goerigk@unimelb.edu.au}
	%\phone{+61 3 834 46784}
	\affiliation{School of Chemistry, The University of Melbourne, Victoria 3010, Australia}

%% The abstract environment will automatically gobble the contents
%% if an abstract is not used by the target journal.
%%%%%%%%%%%%%%%%%%%%%%%%%%%%%%%%%%%%%%%%%%%%%%%%%%%%%%%%%%%%%%%%%%%%%
\begin{abstract}
We continue our work on the long-range corrected double-hybrid density functionals (LC-DHDFs) $\omega$B2PLYP and $\omega$B2GP-PLYP that we developed in the context of time-dependent (TD) Density Functional Theory (DFT) to enable the robust description of singlet-singlet excitations [\textit{J. Chem. Theory Comput.} \textbf{15}, 4735 (2019)]. In our initial study, we only assessed the impact of a long-range correction (LC) on BLYP-based DHDFs, and herein we extend our understanding by providing the first test of PBE-based LC-DHDFs within the established TD-DHDF scheme. Moreover, this study is one of few that provides a direct comparison between TD-DHDFs and their faster Tamm-Dancoff-Approximation variants (TDA-DHDFs). Most importantly, this is the first TDA-DHDF study since Grimme and Neese’s TDA-B2PLYP [\textit{J. Chem. Phys.} \textbf{127}, 154116 (2007)]  and the first work on TD-DHDFs that addresses singlet-triplet excitations. We show how the difference between TD-DHDFs and TDA-DHDFs is often negligible for singlet-singlet excitations, but how one has to apply TDA-DHDFs for triplet excitations. For both excitation types, the LC is beneficial to the BLYP-based DHDFs but detrimental to the PBE-based ones. For local-valence and Rydberg excitations, $\omega$B2PLYP and $\omega$B2GP-PLYP as well as the global DHDF PBE-QIDH can be recommended. If a transition exhibits charge-transfer character, $\omega$B2PLYP and $\omega$B2GP-PLYP should be applied. An analysis of the gaps between the first singlet and triplet excited states of our systems revealed that there is room for further improvements to reach better robustness. Until that goal has been achieved, we recommend $\omega$B2PLYP and $\omega$B2GP-PLYP as some of the currently best TDA-DFT methods.
\end{abstract}
\maketitle

%%%%%%%%%%%%%%%%%%%%%%%%%%%%%%%%%%%%%%%%%%%%%%%%%%%%%%%%%%%%%%%%%%%%%
%% Start the main part of the manuscript here.
%%%%%%%%%%%%%%%%%%%%%%%%%%%%%%%%%%%%%%%%%%%%%%%%%%%%%%%%%%%%%%%%%%%%%
% \input{./sections/introduction.tex}
\section{Introduction}
\label{sec:intro}

Benchmark studies of both new and existing density-functional approximations (DFAs) for excited-state problems constitute an essential contribution to both developers and users alike. They do not only provide the first group with important insights but they allow the second to safely pick DFAs that enable an accurate and reliable description of electronic excited states in applications such as photovoltaics, singlet-fission materials, photosynthesis, photocatalysis, or molecular photo-switches, areas of technological relevance~\cite{Ullrich2012,CT_TDDFT_2017,singlet_fission2015,singlet_fission2017,singlet_fission2018,singlet_fission2019,MeroSwitch,photosynthesys2019} in which electron-transfer processes---the topic of this special issue---play a crucial role. 

This work focuses on double-hybrid density functionals (DHDFs),~\cite{B2PLYP,Goerigk_2014} whose superiority over conventional Density Functional Theory (DFT) approaches has been proven not only for ground-state properties,~\cite{GMTKN30,GS_GOERIGK_DHDFs_2011,GMTKN55,wb97m-2,Nisha2018,DSDBLYP,Kozuch_2011,Kozuch_2013,revDSD2019,JMartin_Empirical2019,DHDF-STO} but also for excitation energies and UV/Vis absorption  spectra.~\cite{GRIMME_2007,Goerigk_Moellmann_2009,Goerigk_2009,Goerigk_2010,Goerigk_2011,exciton_ECD,meo2013,momeni2016,MeroSwitch,Schwabe_2017,wB2PLYPwB2GPPLYP} Previous works have shown that DHDFs even compete with or outperform common wave function-based approaches.\cite{GS_GOERIGK_DHDFs_2011,Nisha2018,Goerigk_2010,Schwabe_2017,wB2PLYPwB2GPPLYP} As defined by Grimme in 2006 for his B2PLYP functional,~\cite{B2PLYP} DHDFs consist of a hybrid-DFA\cite{BHLYP} portion and an additional nonlocal second-order perturbative\cite{MP2_1934} correction. Having been initially formulated for electronic-ground state calculations, Grimme and Neese introduced the time-dependent (TD) extension of  B2PLYP in 2007, which gave access to vertical excitation energies at the DHDF level;~\cite{GRIMME_2007} herein we refer to any methods following that idea as ``TD-DHDFs''. In subsequent years, more excited-state studies on both semi- and ``non-empirical''\cite{NonEmpDHDFrev} DHDFs have been conducted, with the first usually outperforming the latter,\cite{Goerigk_2014,ADAMOJCC2020} something that was later also demonstrated for ground-state properties by Mehta et al. in 2018.~\cite{Nisha2018} While we refer the reader to a recently published detailed review of TD-DHDFs,\cite{TDDHDFaccount} we highlight herein merely some of the advantages of TD-DHDFs, namely a better description of electronic circular dichroism (ECD) 
spectra,~\cite{Goerigk_2009} more accurate vertical singlet-singlet valence excitations in large organic dyes,~\cite{Goerigk_Moellmann_2009,Goerigk_2010,Schwabe_2017} and a more balanced description of the difficult\cite{La_Linear_1,La_Linear_2} first two excited states in polycyclic aromatic hydrocarbons (PAHs).\cite{Goerigk_2011,wB2PLYPwB2GPPLYP} Notably, TD-DHDFs were the only DFAs able to reproduce an exciton-coupled ECD spectrum~\cite{exciton_ECD} of a merocyanine dimer, for which a wave-function electron-correlation method\cite{Rhee2007} had to be used before.\cite{Goerigk_2008} 

The aforementioned TD-DHDF studies focussed on BLYP\cite{B88,LYP,STOLL1989} or PBE\cite{PBE1986} based exchange-correlation expressions and left the nonlocal correlation component of the DHDF untouched, until Schwabe and Goerigk successfully applied spin-component- and spin-opposite-
scaling\cite{scsmp2,SOSMP2_2004,Rhee2007,Lars_SCS_2012} (SCS/SOS) techniques to the latter in 2017.\cite{Schwabe_2017} 

Their remarkable performance notwithstanding, TD-DHDFs were until recently unable to properly describe long-range (LR) excitations, particularly of the charge-transfer (CT) type. This was due to the fact that the BLYP/PBE-based hybrid portion of the DHDFs was of a global nature akin to conventional global-hybrid\cite{BHLYP} DFAs. This problem was addressed by us in 2019 when we presented the first two long-range corrected (LC)---or range-separated---TD-DHDFs (TD-LC-DHDFs) optimized for excited states, namely $\omega$B2PLYP and $\omega$B2GP-PLYP.\cite{wB2PLYPwB2GPPLYP} Based on our findings, these two methods represent the two most accurate and robust DFAs in the description of vertical singlet-singlet excited states of organic molecules with equal error bars for local-valence, Rydberg and CT transitions. In addition, we reported the best TD-DFT results to date for the first two excitations in linear and non-linear PAHs. The improved excitation energies could be traced back to the introduction of the LC scheme, as all other functional components had been kept identical to those of their global parent DHDFs B2PLYP and B2GP-PLYP.\cite{B2GPPLYP}  Note that $\omega$B2PLYP and $\omega$B2GP-PLYP were by no means the first published LC-DHDFs, but that all previous LC-DHDFs had exclusively been developed for ground-state properties,\cite{B2P3LYP,wb97x-2,RSX-QIDH,RSX-0DH} with some following a different definition of the DHDF concept.\cite{lrc-xyg3,wb97m-2,RSDHf,CAM-DH} Moreover, not all of those methods are compatible with the TD-DHDF scheme by Grimme and Neese,\cite{B2P3LYP,lrc-xyg3,wb97m-2} which is why they will not be further considered in the following.

Despite our initial success with $\omega$B2(GP-)PLYP and despite both DFAs having already been used in applications,\cite{Pishchalniko2019,WOLLER2020,ALIPOUR2020,One1eSIEDaleLars} our previous study left some questions open. While we have provided proof that the LC scheme improved upon BLYP-based global TD-DHDFs, we have not yet generalized this finding for the other popular class of ``non-empirical'' TD-DHDFs based on PBE exchange and correlation. In this study we intend to rectify this omission and extend our analysis to recently published PBE-based LC-DHDFs that had been parametrized to reproduce the ground-state energy of the hydrogen atom;\cite{RSX-QIDH,RSX-0DH} note that while this did not make them strictly one-electron self-interaction error free,\cite{One1eSIEDaleLars} some of those methods also performed reasonably well for ground-state energies of one-electron mononuclear systems other than the hydrogen atom,\cite{One1eSIEDaleLars} and showed encouraging results for other many-electron self-interaction-error problems.\cite{RSXvsOT}

In the past, the majority of BLYP-based TD-DHDF studies have been carried out using the full TD-DFT scheme.\cite{Goerigk_2009,Goerigk_Moellmann_2009,Goerigk_2010,Goerigk_2011,exciton_ECD,wB2PLYPwB2GPPLYP} Most studies of PBE-based global DHDFs have exclusively used the Tamm-Dancoff Approximation\cite{TDA-DFT} (TDA-DFT) instead,\cite{meo2013,momeni2016,Sancho-Garcia2016,BREMOND2017,ADAMOJCC2020} with Ref.\citenum{ADAMOIJQC2020} being the only exception we are aware of. There has only been one study that consistently compared BLYP and PBE-based DHDFs within both the TD- and TDA-DFT frameworks.\cite{Schwabe_2017} As TDA-DFT offers a slightly more cost-efficient alternative to TD-DFT for larger chromophores, we intend to close the present knowledge gap and assess LC-DHDFs within the TDA and compare them with the more rigorous TD-DHDF scheme. 

The third and most important question that we intend to answer in the present work is to determine how well TD(A)-LC-DHDFs describe triplet states, in particular singlet-triplet excitations. To date, only one study has ever investigated the TDA-B2PLYP approach for such excitations, namely the original 2007 study by Grimme and Neese.\cite{GRIMME_2007} All subsequent TD-DHDF and TDA-DHDF benchmark studies and applications have been limited to singlet-singlet excitations,\cite{Goerigk_Moellmann_2009,Goerigk_2009,Goerigk_2010,Goerigk_2011,exciton_ECD,meo2013,momeni2016,MeroSwitch,Schwabe_2017,wB2PLYPwB2GPPLYP,vintonyak2010,send2011,prlj2016,alipour2016,BREMOND2017,ADAMOIJQC2020,ADAMOJCC2020} presumably due to the lack of a code that could handle such excitations. We therefore present the first in-detail examination of such excitations for a broader range of global and LC-DHDFs within both the TD-DFT and TDA-DHDF frameworks. Of particular importance will be analyzing the impact of the LC scheme on the description of singlet-triplet excitations. Understanding the role that triplet excitations play is especially important in the development of optoelectronic devices. For instance, photovoltaic devices based on singlet-fission are potential candidates to overcome the Shockley-Queisser limit,\cite{ShockleyQueisser} but the first triplet state (T1) must lie in an appropriate energy window between the ground (S0) and first excited singlet (S1) states.~\cite{singlet_fission2006,singlet_fission2015,singlet_fission2017,singlet_fission2018,singlet_fission2019} In order to assist such developments with calculations, a DFA 
must be reasonably robust and accurate for both singlet and triplet states in addition to having the same accuracy for local-valence and LR excitations. In this study we point out if that challenging requirement can be met by the exisiting global and LC-DHDFs or if future developments and improvements are needed.

The discussion of our results consists of two parts. In the first part, we revisit singlet-singlet excitations on sets that we analyzed in our previous paper\cite{wB2PLYPwB2GPPLYP} and a new set on ``exotic molecules'' that had never been assessed with TD(A)-DFT methods before.\cite{exotic2020} Re-analyzing singlet-singlet transitions has two reasons, the first being a comparison between our previously published data and four PBE-based DHDFs---two of which are LC-DHDFs---and the second being an analysis and comparison between TDA and TD-DFT results. After having given the detailed overview of singlet-singlet transitions we proceed to the second part of our discussion, which consists of an analysis of singlet-triplet transitions based on a variety of different sets, incl. hitherto unpublished reference data for the updated Gordon benchmark set\cite{leang2012,Schwabe_2017} that Schwabe and Goerigk generated in 2017 but had only published for singlet-singlet excitations.\cite{Schwabe_2017} The discussion in our second part comprises the conventional analysis of S0-T$N$ transitions, with $N\geq$1, as well as of S1-T1 gaps. The latter allows us to identify robust methods for future applications such as singlet-fission materials. As such, we hope to provide a thorough overview of the current state of TD(A)-DHDFs with recommendations and guidelines for future developments that should address any of the shortcomings and gaps that we may identify.

\section{Computational details}\label{sec:compdet}
The technicalities of the TD(A)-DFT formalism and how to perform TD(A)-DHDF calculations have been well-documented in the literature\cite{TDA-DFT,TDDFTGrossKohn,CASIDA_TDDFT_1995,CASIDA_2009,TDDFT_BOOK_2012,Ullrich2012,TD-DFT-Book,TD-DFT-Review,GRIMME_2007,Schwabe_2017,wB2PLYPwB2GPPLYP,ADAMOJCC2020} and in our recent account,\cite{TDDHDFaccount} which is why we keep our own explanations brief. Within the adiabatic approximation,\cite{TDDFTGrossKohn,AdiabatApprox} which assumes that  the time-dependent exchange-correlation kernel can be substituted by the time-independent one from ground-state DFT, a linear-response TD-DFT calculation constitutes in solving a non-Hermitian eigenvalue problem that is akin to the random phase approximation (RPA) from wave-function theory:\cite{TDHF-TDHFformolecules}

\begin{equation}\label{RPA}
\left(\begin{array}{cc}
\mathbf{A}&\mathbf{B}\\
\mathbf{B}&\mathbf{A}
\end{array}\right)\left(\begin{array}{c}
\mathbf{X}\\
\mathbf{Y}
\end{array}\right)=\Delta E^{TD-DFT}\left(\begin{array}{cc}
\mathbf{1}&\mathbf{0}\\
\mathbf{0}&\mathbf{-1}
\end{array}\right)\left(\begin{array}{c}
\mathbf{X}\\
\mathbf{Y}
\end{array}\right)\ ,
\end{equation}

where $\mathbf{A}$ and $\mathbf{B}$ are matrices that describe single-particle excitations and corresponding de-excitations, with the individual matrix elements containing information on the occupied and virtual molecular orbitals obtained from the preceding DFT ground-state calculation, as well as the underlying exchange-correlation kernel. $\mathbf{X}$ and $\mathbf{Y}$ are the corresponding eigenvectors and $\Delta E^{TD-DFT}$  is the vertical excitation energy. In the TDA, the elements of $\mathbf{B}$ are essentially set to zero, which results to an equation that is the equivalent of the Configuration Interaction Singles (CIS) problem in wave-function theory:\cite{CIS1}

\begin{equation}\label{TDAEQ}
\mathbf{A}\mathbf{X}=\Delta E^{TDA-DFT}\mathbf{X},
\end{equation}

where $\Delta E^{TDA-DFT}$ is the vertical TDA-DFT excitation energy; for rigorous discussions of TDA-DFT, see e.g. Refs.\citenum{TDA-DFT} or \citenum{TDA_polythiophenes}. A test calculation with the hybrid part of the DHDF B2PLYP for the first 10 singlet excited states of the water molecule showed that the TDA-DFT implementation in ORCA\cite{orcawires,orca4} is about three times faster than the TD-DFT implementation.

According to Grimme and Neese, a TD(A)-DHDF excitation energy is obtained in a two-step fashion, which can be summarized as:\cite{GRIMME_2007} 

\begin{equation}
	\label{eq:DH_energies}
	\Delta E^{TD(A)-DHDF}= \Delta E^{TD(A)-DFT} + a_C \Delta E_C^{CIS(D)}.
\end{equation}

This means that the excitation energy $\Delta E^{TD(A)-DFT}$ of the hybrid part of the DHDF is obtained via one of the two formalisms shown in Eqns. \ref{RPA} and \ref{TDAEQ}, and then corrected by using Head-Gordon's Configuration Interaction Singles with Perturbative Doubles Correction [CIS(D)] approach.\cite{HEADGORDONCISD} The latter gives rise to the correction term $\Delta E_C^{CIS(D)}$ in Eqn. \ref{eq:DH_energies} which is scaled by the parameter $a_C$; this parameter is usually the same as defined for the nonlocal correlation term in ground-state DHDFs.~\cite{B2PLYP} Potential shortcomings and limitations of using the CIS(D) correction in TD-DHDFs have been discussed in Ref. \citenum{TDDHDFaccount}; therein, we also mentioned that the computational effort of the CIS(D) step made up about 2-4 \% of the total time to calculate a single state in the benzene and naphthalene molecules.

Note that ORCA has been the only program so far that has provided general users with the possibility to conduct such calculations. However, it has been often overlooked that initially only global TDA-DHDF calculations were possible. Full TD treatments with global DHDFs have only been enabled in version 4.1, while our LC-DHDFs, namely $\omega$B2(GP-)PLYP, have been implemented into version 4.2. The currently available versions only allow for the calculations of singlet-singlet excitation energies, which is why the herein presented results are based on our local version of the ORCA code into 
which we have implemented the CIS(D) correction for triplet excitations. When dealing with LC-DFAs, the short-range (SR) part of the exchange functional usually needs to be modified. This had already been done for the Becke88 component\cite{CAM-B3LYP} in our published implementation of  $\omega$B2(GP-)PLYP, but for the present study we also implemented the SR-component of PBE exchange\cite{Iikura_2001} to allow the treatment of PBE-based LC-DHDFs. Our modifications will be made available to the general user in the next ORCA release.

The global DHDFs assessed in this work are B2PLYP,\cite{B2PLYP} B2GP-PLYP,\cite{B2GPPLYP} PBE-0DH,\cite{PBE-0DH} and PBE-QIDH.\cite{PBE-QIDH} The impact of an LC scheme on those four DHDFs is analyzed by testing the LC-DHDFs $\omega$B2PLYP,\cite{wB2PLYPwB2GPPLYP} $\omega$B2GP-PLYP,\cite{wB2PLYPwB2GPPLYP} RSX-0DH,\cite{RSX-0DH} and RSX-QIDH.\cite{RSX-QIDH} Note that this 
is the first time that $\omega$B2PLYP and $\omega$B2GP-PLYP are tested within the TDA-DHDF scheme, while it is the first time that the ``RSX'' methods are assessed for the calculation of excitation energies within both the TDA and TD-DHDF schemes. We tested our ``RSX'' implementation by reproducing RSX-QIDH's total energy for the H atom and the mean absolute deviation for the SIE4x4\cite{GMTKN55} test set, as reported in Ref. \citenum{RSX-QIDH}; we only observed a marginal difference of 0.08 kcal/mol for the latter most likely due to differences in program settings and the lack of the resolution-of-the-identity for the nonlocal correction in the ``RSX'' developers' program code. 

There is substantial evidence in the literature that TD(A)-DFT for DFAs below the hybrid level are not suitable,\cite{Thiel2008_2,Goerigk_Moellmann_2009,jacquemin2009,leang2012,TDDHDFaccount}  which is why we refrain from testing them. It has also been established that DHDFs on average outperform hybrid functionals,\cite{Goerigk_Moellmann_2009,TDDHDFaccount} which is why we only test a limited selection of hybrid functionals either due to their popularity [B3LYP~\cite{B3LYP,b3lyp_2} and BHLYP~\cite{BHLYP}] or because they have been LR-corrected and are thus the closest competitors of LC-DHDFs [CAM-B3LYP~\cite{CAM-B3LYP} and $\omega$B97X~\cite{wb97x}]. Note that the same hybrids were chosen for our previous work on $\omega$B2(GP-)PLYP; in order to maintain consistency we included them again in the present work. The SCF convergence criterion was set to 10$^{-7}E_h$ (10$^{-8}E_h$ for the updated Gordon set) along with ORCA's numerical quadrature grid ''4'' (grid ''5'' and finalgrid ''6'' for the updated  Gordon set). The resolution-of-the-identity technique was used with appropriate auxiliary basis sets in the perturbative steps.\cite{RIapprox,aux_basis_MP2}

Our analysis of singlet-singlet excitations comprises a set that we initially used as training set\cite{Schwabe_2017} for $\omega$B2PLYP and $\omega$B2GP-PLYP,\cite{wB2PLYPwB2GPPLYP} the Gordon benchmark set,\cite{leang2012} which had been updated by Schwabe and Goerigk in 2017,\cite{Schwabe_2017} as well as our collection of CT excitations for many of which we presented updated references values in 2019.\cite{wB2PLYPwB2GPPLYP} We took TD-DFT results for B2(GP-)PLYP, their LC versions, B3LYP, BHLYP, CAM-B3LYP and $\omega$B97X from our previous study to compare them with the new TD-DFT results that we obtained herein for PBE0-DH, PBE-QIDH functionals and their LC versions. New TDA-DFT numbers for those sets were calculated for all DFAs for this work.  In addition, we also include an analysis of a fairly new set proposed by Loos, Jacquemin, and co-workers that contains some ''exotic'' molecules,~\cite{exotic2020} for which no TD(A)-DFT assessment has been performed until now. 

Our analysis of singlet-triplet excitation energies starts with a small set used by 
Grimme and Neese for their initial work on TDA-B2PLYP.\cite{GRIMME_2007} We then provide an analysis of the aforementioned updated Gordon benchmark set with hitherto unpublished reference values that were obtained as part of Schwabe and Goerigk's singlet-singlet excitation study.\cite{Schwabe_2017} This is followed by an analysis of those molecules belonging to Thiel's famous set\cite{Thiel2008} that had not been included in any of the previous sets. We then finalise our analysis with two sets published by Loos, Jacquemin, and co-workers, namely one containing difficult cases published in 2018\cite{loos2018} as well as the aforementioned 
``exotic molecules'' set from 2020.\cite{exotic2020} For the 2018 set we calculated one additional singlet-triplet CT reference energy using Dalton 2018.2\cite{daltonpaper,dalton2} (see Section \ref{sec:loos} for more details). All  molecular geometries were taken from the literature and all calculations were carried out with the same atomic-orbital (AO) basis sets as used as in the original benchmark studies.\cite{Schwabe_2017,exotic2020,Thiel2008,loos2018,wB2PLYPwB2GPPLYP} More details on the nature of each set, the reference values, and the AO basis sets is provided in each section. We refrain from a basis-set dependence study as such studies have been done for both the reference values as well as TD-DHDFs before.\cite{Schwabe_2017,exotic2020,Thiel2008,loos2018,wB2PLYPwB2GPPLYP,TDDHDFaccount} Note that previously it was established that even for full CIS(D) ($a_C=1$)  local-valence and Rydberg excitation energies only differed by about 0.03 eV when going from a triple to a large quadruple-$\zeta$ basis.\cite{Schwabe_2017} Unsurprisingly, this difference was lower for DHDFs with decreasing $a_C$, for instance only 0.01 eV for TD-B2PLYP ($a_C=0.27$) and TD-PBE0-DH ($a_C=0.125$ ).\cite{Schwabe_2017}

For most molecules, several excitation energies from the S0 to various triplet (T$N$, with $N\geq$1) states are calculated and our analyses comprise statistics over all those excitations. In addition, we also provide separate analyses over the splitting between S1 and T1 states, a quantity not often analyzed in benchmark studies but crucial for applications such as singlet-fission, for which robust methods need to be found that describe low-lying singlet and triplet states equally well. 

All our analyses comprise standard statistical indicators such as mean deviations (MDs), mean absolute deviations (MADs), root-mean-square deviations (RMSDs) and error ranges ( $\Delta_{\text{err}}$). All deviations have been calculated as the difference between the assessed method and the reference value, which means that a negative deviation automatically implies an underestimation of an excitation energy or S1-T1 gap. While our discussion below is restricted to only some of those statistical values, all results are listed in the Supplementary Material.

In the context of our discussion, it is also important to consider the often-cited ``chemical accuracy'' goal in excited-state benchmarking. It is an arbitrarily defined value and some studies have suggested a value of 0.05 eV,\cite{SendFurcheAdEx2011,LoosJacqueminChemAcc2018} while others prefer to already regard an accuracy of 0.1 eV as a success.\cite{exrev,Goerigk_Moellmann_2009} Our study can be interpreted within the context of either of these two definitions. 

\section{Revisiting vertical singlet-singlet excitations}
\label{sec:singlets}

\subsection{The  $\omega$B2(GP-)PLYP training set}
\label{sec:training}
\begin{figure}
	\centering
	\includegraphics[width=1\linewidth]{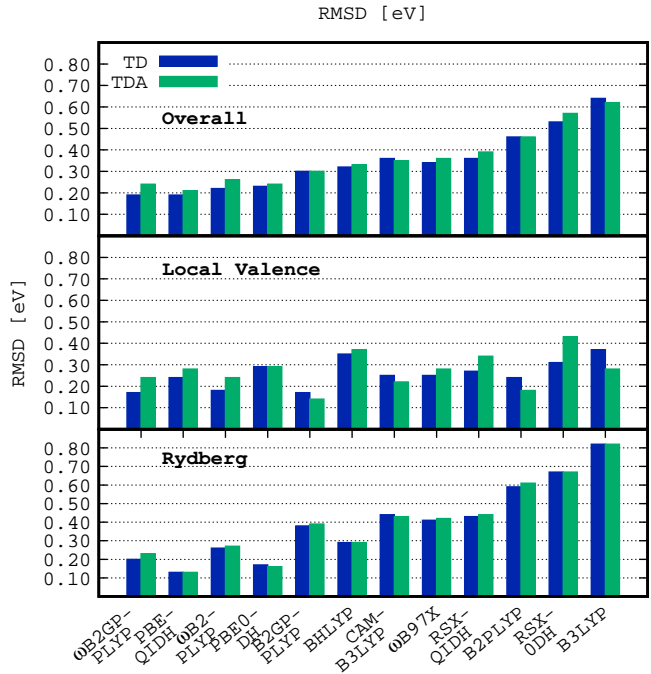}
	\caption{RMSDs (in eV) for all, local-valence, and Rydberg excitations in the  $\omega$B2(GP-)PLYP training set. TD-DFT results are shown in blue, and TDA-DFT results in green. The aug-cc-pVTZ AO basis was used in all cases.}
	\label{fig:fitsettdtda}
\end{figure}

Our ``training set'', which is a slight modification of a set presented by Schwabe and Goerigk in 2017\cite{Schwabe_2017} and was used in our previous work in order to determine the values of the LR parameters in \mbox{$\omega$B2(GP-)PLYP},\cite{wB2PLYPwB2GPPLYP} contains 34 excitations divided into 28 valence and 6 Rydberg vertical singlet-singlet excitations, where the latter were weighted by a factor of 5 in order to achieve an approximate 1:1 ratio. Reference values are based on Coupled Cluster Singles Doubles with Perturbative Triples Excitation Correction\cite{ccsdr3} [CCSDR(3)]  and the aug-cc-pVTZ\cite{Dunning_1989,Kendall_1992} AO basis set.\cite{Schwabe_2017}  As already mentioned, we return to this set in order to assess new DFAs and also the TDA variants, since neither of them had been previously considered; as such it is not used for training new approaches. We present the RMSD values for both TD and TDA approaches in Figure \ref{fig:fitsettdtda} where the values are shown for the entire set and also broken down into local-valence and Rydberg transitions; the methods are ordered according to their TD values for the entire set. All excitation energies and statistical data are provided in Tables S1 and S2 in the Supplementary Material. Since most of these DFAs studied herein were discussed in our previous work,\cite{wB2PLYPwB2GPPLYP} we focus on the TDA results and only comment on the TD-DFT values where relevant. 

Firstly, we notice that $\omega$B2GP-PLYP, PBE-QIDH,  $\omega$B2PLYP, and PBE0-DH position themselves as the best of the tested DFAs for both the TD and TDA variants for the entire set. The RMSDs are very similar and range from 0.19 to 0.23 eV (TD) and from 0.21 to 0.26 eV (TDA). Secondly, most tested DFAs do not show any significant difference between the TD and TDA algorithms with an average RMSD difference between the two schemes being about $\pm$0.02~eV for the whole set. However, we see that if differences between TD and TDA occur, they tend to be more pronounced for the local-valence category with the five Rydberg excitations being less affected by the change in algorithm, as we will see further below. For the cases with noticeable differences between the TDA and TD algorithms, we see in Tables S1 and S2 that the TDA generally induces a blueshift, which is known from previous studies on lower-rung DFAs\cite{TDA-DFT,TDA_polythiophenes,moreonTDA} as well as global DHDFs.\cite{Schwabe_2017} Ref. \citenum{TDA_polythiophenes} provides a detailed analysis of this finding for polythiophenes and shows how the difference between TD and TDA-DFT decreases with larger chromophore size. This has been explained with the fact that the TDA neglects resonant-antiresonant coupling, which is usually governed by the exchange component of the electron-hole interaction. The authors of Ref. \citenum{TDA_polythiophenes}  have argued how the resulting deviation between TDA-DFT  and TD-DFT for smaller systems seems to be connected to the spatial extent of electron and hole wave functions.

Among the TDA-DFT results, PBE-QIDH has the best RMSD for the entire set with a value of 0.21~eV, followed by $\omega$B2GP-PLYP and PBE0-DH with the same value of 0.24~eV, and $\omega$B2PLYP with 0.26~eV, respectively. One cannot but notice that B2GP-PLYP also performs very well with an RMSD of 0.30~eV, which is identical for both algorithms. The very good results for both PBE-QIDH and B2GP-PLYP are most likely due to their relatively high amounts of Fock exchange---69 and 65~\%, respectively---which seems to compensate for the absence of an LC. BHLYP follows B2GP-PLYP with an RMSD of 0.33~eV.  $\omega$B97X, CAM-B3LYP, and RSX-QIDH share similar RMSDs of 0.36, 0.35, and 0.39~eV, respectively. Finally, the remaining three DFAs (B2PLYP, RSX-0DH, and B3LYP) are the ones with the worst overall results with RMSDs of 0.46, 0.57, and 0.62~eV.

When using TDA-DFT for local-valence excitations only, B2GP-PLYP displays the best result with an RMSD of 0.14~eV. Note that it is also the best when using the TD scheme along with TD-$\omega$B2GP-PLYP (RMSD = 0.17~eV). B2GP-PLYP is followed by B2PLYP, with an RMSD of 0.18~eV, a considerable improvement over its TD result of 0.24~eV. CAM-B3LYP follows with a value of 0.22~eV, slightly better compared to $\omega$B2GP-PLYP and $\omega$B2PLYP, which share the same result of RMSD = 0.24~eV. We also see that PBE-QIDH, B3LYP, and $\omega$B97X perform equally with RMSD = 0.28~eV, closely followed by PBE0-DH with RMSD = 0.29~eV. The remaining DFAs have RMSDs above 0.3~eV: RSX-QIDH (0.34~eV), BHLYP (0.37~eV), and RSX-0DH (0.43~eV). Note that TDA-RSX-0DH performs worse than TD-RSX-0DH with an increase of 0.12 eV in the RMSD. The TDA results show a noticeable difference to the TD ones for \text{$\omega$B2(GP-)PLYP}, for which we previously reported improvements compared to their global parent DHDFs due to the LC scheme for B2PLYP and the same RMSDs for B2GP-PLYP and its LC version.\cite{wB2PLYPwB2GPPLYP} We attribute the difference between the TD and TDA results to the fact that \text{$\omega$B2(GP-)PLYP} had been optimized for the TD algorithm on this particular set. Further cross-validation in the following sections will show if we observe a similar trend.

The results for the Rydberg category reveal remarkable performance for PBE-QIDH for both the TDA and TD schemes (RMSD = 0.13~eV in both cases). When focussing solely on the TDA results, PBE0-DH follows PBE-QIDH with a value of 0.16~eV. $\omega$B2GP-PLYP (RMSD = 0.23~eV), $\omega$B2PLYP (RMSD = 0.27~eV) and BHLYP (RMSD = 0.29~eV) are the remaining three TDA-DFT approaches with values below 0.3~eV. B2GP-PLYP drops four places in the ranking compared to the local-valence category, with a large RMSD of 0.39~eV. Surprisingly, $\omega$B97X (0.42~eV) performs poorly in this category despite being an LC-hybrid, with a clear tendency to overestimate these energies (see Supplementary Material for more details). All the remaining methods, regardless of whether they tend to be global or LC-DFAs, display larger RMSDs that range from 0.43~eV (CAM-B3LYP) to 0.82~eV (B3LYP). We see exactly the same behavior for the TD results.

The effect of the LC for this set seems to depend strongly on the underlying exchange-correlation DFA but interestingly also on the type of algorithm. The two BLYP-based DHDFs clearly benefit from the LC for the Rydberg excitations for both the TDA and TD versions, which was our intention when developing them. Improvements can also be found for local-valence excitations for TD-$\omega$B2PLYP, while the statistics remain unchanged for TD-$\omega$B2GP-PLYP. However, a slight worsening of the results  (between 0.05 and 0.1 eV) is observed by introducing the LC to the TDA variants. In the case of the PBE-based DHDFs, we see that for local-valence excitations the LC makes them worse compared to their global versions for both algorithms: TD(A)-PBE-QIDH = 0.24(0.28)~eV vs TD(A)-RSX-QIDH = 0.27(0.34)~eV and TD(A)PBE0-DH = 0.29(0.29)~eV vs TD(A)-RSX-0DH = 0.31(0.43)~eV. However, when comparing TDA-LC-DHDFs with TD-LC-DHDFs for Rydberg excitations, the differences in the excitation energies seem to be negligible regardless of the underlying exchange-correlation DFA. For instance, TDA-$\omega$B2PLYP has an RMSD that is by only 0.01 eV higher than for its TD version, while TDA-$\omega$B2GP-PLYP shows a slightly larger increase compared to its TD version (by 0.03~eV). For PBE-based DFAs we see the same behavior, where TDA-RSX-QIDH's RMSD has increased by 0.01~eV and RSX-0DH remains the same for both algorithms. We will see later if the same trend can also be observed for CT transitions.

\subsection{The updated Gordon benchmark set}
\label{sec:gordon_sing}

The original benchmark set by Gordon and co-workers~\cite{leang2012} comprises a total of 63 singlet-singlet electronic excitations made up of 33 valence and 30 Rydberg states from 14 different molecules. The geometries were later reoptimized by Schwabe and Goerigk in 2017,~\cite{Schwabe_2017} where they also provided new full and estimated Approximate
Coupled-Cluster Singles Doubles Triples data (CC3~\cite{Koch1997}/aug-cc-pVTZ), which replaced the original experimental references; for a discussion on the benefit of using wave-function over experimental references in benchmarking electronic-structure methods, see e.g. Refs. \citenum{Schwabe_2017}, \citenum{TDDHDFaccount} and \citenum{Goerigk_Mehta2019}. All results and statistical values are shown in Tables S3 and S4.

\begin{figure}
	\centering
	\includegraphics[width=1\linewidth]{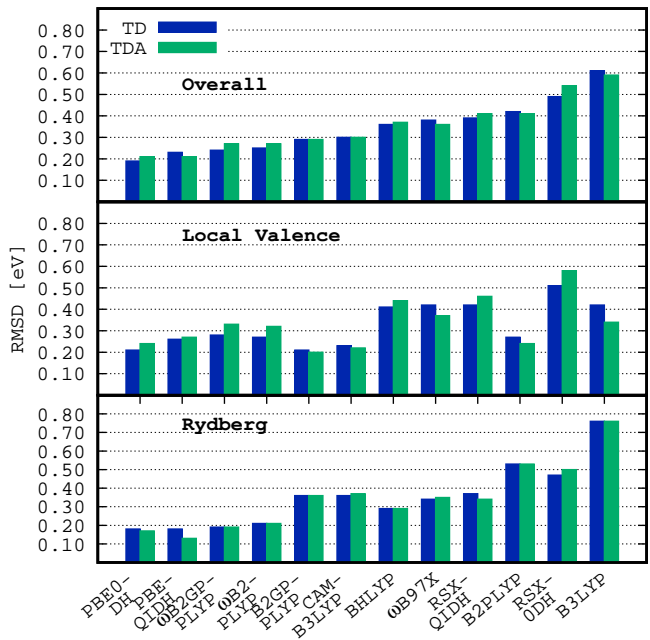}
	\caption{RMSDs (in eV) for all, local-valence, and Rydberg states in the updated Gordon benchmark set. TD-DFT results are shown in blue, and TDA-DFT results in green. The aug-cc-pVTZ AO basis was used in all cases.}
	\label{fig:gordon_sing}
\end{figure}

The results are compiled in Figure~\ref{fig:gordon_sing} and the trend is very similar to the one presented in the previous section. More specifically, PBE0-DH, PBE-QIDH and $\omega$B2GP-PLYP display the lowest overall RMSD values of 0.21/0.19, 0.21/0.23 and 0.27/0.24~eV for TDA/TD, respectively, closely followed by $\omega$B2PLYP (0.27/0.25~eV for TDA/TD). Once again, B2GP-PLYP shows the same result for both algorithms with RMSD = 0.29~eV. Meanwhile, CAM-B3LYP displays an improvement compared to the last set with RMSD =  0.30~eV for both TD and TDA, 0.06/0.05~eV lower for TD/TDA than in the previous set. $\omega$B97X and BHLYP perform similarly with values ranging from 0.35 to 0.38~eV, depending on the chosen method. RSX-QIDH performs slightly better compared to B2PLYP when using a TD algorithm (RMSD = 0.39 and RMSD = 0.42~eV, respectively) but performs equally within the TDA (RMSD = 0.41~eV). Finally, RSX-0DH and B3LYP rank once again in the last two positions with RMSDs of 0.5~eV and beyond.

When switching to a TDA algorithm, the RMSDs increase for the local-valence category compared to the full TD-DFT framework, but the general trends remain the same as discussed in the previous section with some specific exceptions. For instance, while TDA-B3LYP and TDA-$\omega$B97X show an improvement of 0.08 and 0.05~eV in this category, TDA-RSX-0DH, TDA-$\omega$B2GP-PLYP, and TDA-RSX-QIDH's RMSDs deteriorate by 0.08~eV for the first and 0.05~eV for the last two when compared with the TD-DFT algorithm. For the TDA-DFT formalism, B2GP-PLYP is the best DFA for local-valence transitions with an RMSD of 0.20~eV, followed by CAM-B3LYP (RMSD = 0.22~eV), PBE0-DH (RMSD = 0.24~eV), and B2PLYP (RMSD = 0.24~eV). PBE-QIDH moves down one position in the ranking compared to its TD version with a value of 0.27~eV, being the last DFA with a value lower than 0.3~eV. $\omega$B2PLYP, the closely related $\omega$B2GP-PLYP, and B3LYP share similar results of 0.32, 0.33, and 0.34~eV, respectively. Finally, $\omega$B97X has a value of 0.37~eV, and every other DFA has an even larger RMSD, for instance RMSD(RSX-QIDH) =  0.46 ~eV and RMSD(RSX-0DH) = 0.58~eV. 

In contrast to the local-valence excitations, the trends for Rydberg states are similar to the ones described for the ``training'' set in the previous section and we do not see any significant changes that are worth being discussed in detail.

\begin{table*}
	\centering
	\footnotesize
	\setlength\tabcolsep{2pt}
	\caption{\label{tab:exotic1} Statistical results (in eV) for the ``exotic-molecules'' set. The results are presented for TD-DFT and TDA-DFT with the latter shown in parentheses.$^a$}
	%	\begin{tabular}{\textwidth}{r*{6}{D{.}{.}{2}}} %8l + 1D
	\begin{tabular}{p{2.25cm}p{2.25cm}p{2.25cm}p{2.25cm}p{2.25cm}p{2.25cm}r*{6}{D{.}{.}{2}}} %8l + 1D
		\hline
		\multicolumn{1}{c}{\centering }&\multicolumn{1}{c}{\centering B3LYP} & \multicolumn{1}{c}{\centering BHLYP} & \multicolumn{1}{c}{\centering CAM-B3LYP} & \multicolumn{1}{c}{\centering $\omega$B97X} & \multicolumn{1}{c}{\centering B2PLYP} & \multicolumn{1}{c}{\centering B2GP-PLYP} \\
		\hline	
		\multicolumn{1}{c}{\centering  MD}                       &$-$0.22($-$0.12)&$-$0.04(0.04)&$-$0.17($-$0.08)&$-$0.11($-$0.01)&$-$0.09($-$0.06)&$-$0.03($-$0.01)\\
		\multicolumn{1}{c}{\centering  MAD}                      &0.22(0.16)&0.19(0.18)&0.19(0.14)&0.16(0.13)&0.13(0.10)&0.15(0.08)\\
		\multicolumn{1}{c}{\centering  RMSD}                     &0.30(0.26)&0.29(0.26)&0.26(0.22)&0.22(0.20)&0.16(0.15)&0.10(0.10)\\
		\multicolumn{1}{c}{\centering  $\Delta_{\text{err}}$}    &0.95(0.98)&1.31(1.12)&0.69(0.73)&0.81(0.76)&0.49(0.61)&0.86(0.42)\\
		\hline
	\end{tabular}
	\begin{tabular}{p{2.25cm}p{2.25cm}p{2.25cm}p{2.25cm}p{2.25cm}p{2.25cm}r*{6}{D{.}{.}{2}}} %8l + 1D
		%\begin{tabularx}{\textwidth}{|p{1in}|p{5.75in}|l*{9}{D{.}{.}{2}}} %8l + 1D
		\hline
		\multicolumn{1}{c}{\centering }& \multicolumn{1}{c}{\centering $\omega$B2PLYP}& \multicolumn{1}{c}{\centering $\omega$B2GP-PLYP}&\multicolumn{1}{c}{\centering RSX-QIDH}& \multicolumn{1}{c}{\centering RSX-0DH}&\multicolumn{1}{c}{\centering PBE-QIDH}&\multicolumn{1}{c}{\centering PBE0-DH}\\
		\hline	
		\multicolumn{1}{c}{\centering  MD}                       &$-$0.05($-$0.01)&0.00(0.02)&0.05(0.09)&0.03(0.05)&$-$0.11(0.05)&$-$0.11(0.00)\\
		\multicolumn{1}{c}{\centering  MAD}                      &0.17(0.11)&0.19(0.10)&0.21(0.17)&0.24(0.18)&0.17(0.11)&0.17(0.09)\\
		\multicolumn{1}{c}{\centering  RMSD}                     &0.24(0.15)&0.25(0.14)&0.28(0.25)&0.32(0.25)&0.22(0.15)&0.21(0.13)\\
		\multicolumn{1}{c}{\centering  $\Delta_{\text{err}}$}    &1.05(0.58)&1.11(0.51)&1.25(1.17)&1.45(1.03)&0.84(0.66)&0.84(0.53)\\
		\hline
	\end{tabular}

	$^a$~All calculations were performed with the aug-cc-pVTZ AO basis set.
\end{table*}

We also see the same trends as for the ``training set'' when comparing BLYP and PBE-based DHDFs in regard to the introduction of an LC. For instance, in the local-valence TDA algorithm, B2GP-PLYP presents an RMSD of 0.20~eV that increases to 0.33~eV when using $\omega$B2GP-PLYP. Similarly, B2PLYP also shows an increase from 0.24 to 0.32~eV due to the inclusion of LC. In a similar way, but with notably larger energy differences, PBE-based DHDFs show the same behavior. PBE-QIDH and RSX-QIDH have RMSDs of 0.27 and 0.46~eV, respectively, and PBE0-DH's RMSDs worsens from 0.24 to 0.58~eV due to the LC. Regarding the TDA-Rydberg excitations, once again BLYP-based DHDFs benefit from the inclusion of the LC scheme while PBE-based DHDFs go in the opposite direction. For example, B2PLYP performs poorly for this kind of excitations (RMSD = 0.53 eV) which is greatly improved when using $\omega$B2PLYP (RMSD = 0.21 eV). B2GP-PLYP and $\omega$B2GP-PLYP, in turn, also show a reduction of 0.17~eV due to the LC, lowering their RMSDs from 0.36 to 0.19~eV, respectively. Contrary to this,  PBE-QIDH is the best DFA in this category with an RMSD of just 0.13~eV but, due to the inclusion of the LC scheme, this value raises considerably  to 0.34~eV, almost matching the RMSD obtained for B2GP-PLYP. The same holds for PBE0-DH, whose RMSD rises from 0.17 to 0.50~eV when the LC scheme is applied. We have therefore provided a cross-validation of the trends discussed in the previous section. This shows that the improvement observed for the BLYP-based variants when introducing the LC in the previous section was not a mere consequence of having used that set as a training set for those methods, but it is in fact a tangible trend that can be reproduced.

\subsection{The ``exotic-molecules'' set}
\label{sec:exotic_sing}

Very recently, Loos, Jacquemin, and co-workers presented a benchmark set on ``exotic molecules'', a phrase that implied that each molecule includes (at least) one of the following atoms: F, Cl, Si, or P,~\cite{exotic2020} which are uncommon in usual organic-molecule test sets. Overall, the set comprises 19 local-valence singlet-singlet excitations for 14 different systems with reliable CC3/aug-cc-pVTZ reference values. This work is the first assessing DFT methods with this set. Individual results for each system are shown in Tables S5 and S6.

Based on the RMSDs shown in Table~\ref{tab:exotic1} and contrary to the previous sets, the TDA clearly outperforms the full TD-DFT variants for every tested method by at least 0.01~eV (B2PLYP) and up to 0.11~eV ($\omega$B2GP-PLYP), which might be a 
consequence of the inclusion of the ''exotic'' atoms. When using the TDA algorithm, B2GP-PLYP yields the lowest RMSD value of only 0.10~eV, followed by PBE0-DH (RMSD = 0.13~eV) and its LR-corrected ``child'' $\omega$B2GP-PLYP (RMSD = 0.14~eV). 
The next-best DFAs are ($\omega$)B2PLYP and PBE-QIDH all with RMSD = 0.15~eV. Those six methods are the only ones with RMSDs below 0.2~eV. $\omega$B97X and CAM-B3LYP follow with a values of 0.20~eV and 0.22~eV, respectively. Both PBE-
based LC-DHDFs share the same RMSD of 0.25~eV, followed by BHLYP and B3LYP, which have a value of 0.26~eV in common.

Note that the RMSD is not the only statistical value that improves when moving from a full TD to a TDA-based treatment, but also every other statistical value shown in Table~\ref{tab:exotic1} improves. This includes the MDs, which is a trend that indicates that the TDA variants allow a more robust description of the different excitations. Again, we focus our discussion only on the TDA results. The MDs for half of the presented DFAs seem to be well balanced, as they lie very close to the ideal value of 0~eV. For instance, PBE0-DH's MD is exactly 0.00 eV, $\omega$B97X, B2GP-PLYP and $\omega$B2PLYP exhibit a negligible red-shift of only 0.01~eV, and $\omega$B2GP-PLYP yields an MD of only 0.02~eV. The remaining DFAs have higher absolute mean deviations ranging from 0.04-0.12~eV. The MADs show 
a very similar ranking compared to the RMSDs.

The LC induces a systematic blueshift for all four DHDFs. Contrary to the previously discussed test sets, all TD-LC-DHDFs show larger RMSDs and error ranges compared to their global counterparts. TDA-$\omega$B2PLYP has the same RMSD as B2PLYP and a marginally lower error range. The increase in RMSD and error range caused by the LC is lower for the TDA variant of $\omega$B2GP-PLYP than for its TD version with increases of 0.04 and 0.09~eV, respectively. Stronger increases are observed for the  PBE-based DFAs, for instance an increase of 0.1 eV in the RMSD and of 0.59 eV in the error range for the QIDH-type DFAs.

\subsection{An update on charge-transfer excitations}
\label{sec:CTs}

In the previous three sections, we confirmed again that TD-$\omega$B2PLYP and TD-$\omega$B2GP-PLYP belonged to the top-3 DFAs. We also found that TD-PBE-QIDH and TD-PBE0-DH were some of the best DFAs despite not being LR-corrected. The finding for TD-PBE-QIDH is particularly surprising for Rydberg excitations. Before we continue with a discussion of triplet excitations, it is worthwhile to analyze PBE-QIDH more closely by revisiting come of the CT transitions explored in 
our previous work where we presented newly updated high-level reference data.\cite{wB2PLYPwB2GPPLYP} The original set comprised 17 CT excitations in 9 different systems. However, after closer inspection, we decided to exclude the ``tripeptide'', as the reference value was only based on the Spin-Component-Scaled Approximate Coupled Cluster Singles Doubles\cite{scssoscc2} (SCS-CC2) model. We also excluded two CT excitations of the $n$-$\pi^*$ type in the ``dipeptide'' and ``$\beta$-peptide'' due to their high double-excitation character [27.23\% and 26.41\% at the CCSDR(3) level, respectively]. Therefore, we present a new analysis of eight different systems comprising nine CT excitations. Reference values are based either full CC3, estimated CCSDR(3), estimated EOM-CCSD(T) or---in one case---theoretically back-corrected experimental reference values.\cite{Goerigk_2010,wB2PLYPwB2GPPLYP,loos2018,Moore2015}

\begin{figure}[t]
	\centering
	\includegraphics[width=1\linewidth]{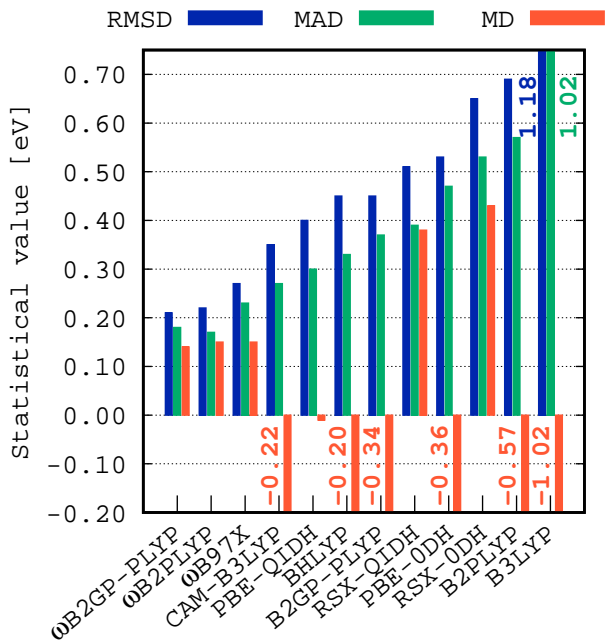}
	\caption{Statistical results (in eV) for the CT test set using TD-DFT. Detailed information about the basis set used on each system can be found in the Supplementary Material.}
	\label{fig:ct_sing}
\end{figure}

MDs, MADs and RMSDs are shown in Figure \ref{fig:ct_sing}, while all excitations and error ranges are shown in Table S7. As expected, LC-DFAs rank on top leading with $\omega$B2GP-PLYP with an RMSD of 0.21~eV, closely followed by $\omega$B2PLYP (0.22~eV). In our previous assessment, we saw that the latter was, in fact, the best DFA for CT. However, with the exclusion of the aforementioned excitations, the RMSD for $\omega$B2GP-PLYP decreased making it the best DFA.  $\omega$B97X and CAM-B3LYP follow with values of 0.27 and 0.35~eV, respectively. PBE-QIDH displays a value of 0.40~eV. Unsurprisingly, the other global hybrids and DHDFs are not useful for describing CT excitations with RMSDs ranging from 0.45~eV (BHLYP and B2GP-PLYP) to  1.18~eV (B3LYP). However, it is surprising that the two PBE-based LC-DHDFs show large RMSDs of 0.51~eV (RSX-QIDH) and 0.65~eV (RSX-0DH). This shows that they cannot compete with the BLYP-based LC-DHDFs.

As expected, most global DFAs severely underestimate  CT excitation energies, e.g. B3LYP with an MD of $-$1.02~eV,  B2PLYP with MD =  $-$0.57 eV or  B2GP-PLYP with MD = $-$0.34~eV. 
Due to the LC, most LC-DFAs systematically overestimate CT transitions, which can be seen from the fact that MADs and MDs have very similar values, for instance MD = 0.14~eV and MAD = 0.18~eV for $\omega$B2GP-PLYP. $\omega$B2PLYP and $\omega$B97X also present with a systematic description of CT excitations; the former has an MD of 0.15~eV and an MAD of 0.17~eV, while the latter has the same MD and an MAD of 0.23~eV. CAM-B3LYP displays a systematic underestimation (MD = $-$0.22~eV and MAD = 0.27~eV), mostly due to the fact that its does not establish the correct asymptotic limit in the LR regime contrary to all other tested LC-DFAs, which establish a fraction of 100\% Fock exchange in that regime. 

The PBE-based DHDFs form a peculiar exception to the above described. While PBE-QIDH has the best MD with a value of only $-$0.01~eV, its MAD of 0.30~eV indicates that this nearly-perfect MD is due to error compensation. Indeed, deviations range from $-$0.72 to 0.83~eV whereas the deviations for $\omega$B2GP-PLYP range from $-$0.17 to 0.42~eV. The introduction of an LC to PBE-QIDH leads to a severe overestimation of the excitation energies with MD(RSX-QIDH) = 0.38~eV. We also see that PBE0-DH already overestimates the excitation energies and that range-separation makes the final outcome worse, with MD(RSX-0DH) = 0.43~eV.  

Finally, in Table S7 we present the error range for each DFA. We observe that the inclusion of the LC-scheme in $\omega$B2(GP-)PLYP is beneficial, as the values have more than halved compared to B2(GP-)PLYP.  Four DFAs yield error ranges below 1~eV; namely $\omega$B2PLYP and $\omega$B2GP-PLYP with values of 0.57 and 0.59~eV, respectively, followed by $\omega$B97X ($\Delta_{\text{err}}$ = 0.78~eV) and CAM-B3LYP ($\Delta_{\text{err}}$ = 0.97~eV). Contrary to that, we observe larger error ranges for PBE-based DHDFs. In the case of PBE0-DH, the value increases from 1.49 to 1.86~eV upon introducing the LC.The value for PBE-QIDH improves from 1.54~eV to 1.26~eV when introducing the LC.

Despite its remarkable performance for the first three test sets, we have shown that PBE-QIDH is unable to describe CT transitions, making it an overall less robust method. We therefore do not agree with the recently made statement that inclusion of the CIS(D) correction alone makes DHDFs a good alternative to LC-hybrids for CT excitations.\cite{ADAMOJCC2020} Note that that statement was made after an analysis of global TDA-DHDFs, without any of the at that time already existing TD(A)-LC-DHDFs, and despite large errors for methods such as TDA-PBE-QIDH that are far from the desired chemical-accuracy threshold---sometimes up to 0.4~eV---and a systematic underestimation of CT transitions in aryl-tetracyanoethylene complexes.\cite{ADAMOJCC2020} Based on our findings, and  contrary to BLYP-based DHDFs, the introduction of the LC makes the PBE-based DHDFs worse, which shows that such a correction alone is no panacea for solving the CT problem. An extension of our present analysis of TD-LC-DHDFs for the sets discussed in Ref.\citenum{ADAMOJCC2020} warrants a separate, future study. Coming back to one or our main aims stated in the Introduction, we proceed in Section \ref{sec:triplets} with an investigation singlet-triplet excitations to determine if similar or different trends are observed compared to Section \ref{sec:singlets}.

\section{Assessment of vertical electronic excitations involving triplet states}
\label{sec:triplets}

\subsection{Connection with early work done by Grimme and Neese}
\label{sec:grimme_neese}

After having obtained promising results for vertical singlet-singlet excitations, we now investigate if global and LC-corrected DHDFs also provide a balanced description for vertical singlet-triplet excitations. As mentioned earlier, the only study to date analyzing such excitations for DHDFs was done for TDA-B2PLYP in 2007,~\cite{GRIMME_2007} when Grimme and Neese examined 22 excitations in seven systems based on experimental data. It therefore makes sense to begin this part of our discussion with an extended analysis of TDA-DHDFs for the same set before proceeding with more detailed investigations.

\begin{table*}[t]
	\caption{\label{tab:Grimme_Neese}Comparison of calculated and experimental vertical singlet-triplet excitation energies (in eV) using TDA-DFT.}
	%	\caption{\label{tab:Grimme_Neese} Comparison of calculated and experimental vertical singlet-triplet excitation energies (in eV) using TDA-DFT.~\footnotemark[1]}
	\begin{ruledtabular}
		\footnotesize
	\begin{tabular}{cccccccccccccc}
	System &  State & Expt.$^b$ & B3LYP  &  CAM  &  $\omega$B97X  &  B2- &  $\omega$B2- &  B2GP-  &  $\omega$B2GP- & RSX- &  RSX- & PBE- & PBE0- \\
	&&&&B3LYP&&PLYP&PLYP&PLYP&PLYP&QIDH&0DH&QIDH&DH\\
	\hline
	%Begin TOTAL (dont delete!)
	\multirow{4}{*}{CO}
	& $^3\Pi$          &6.32&6.07&6.12&6.37&6.26&6.28&6.31&6.31&6.24&6.13&6.12&6.02 \\
	&  $^3\Sigma^+$    &8.51&8.32&8.36&8.45&8.46&8.49&8.50&8.52&8.48&8.40&8.30&8.19  \\
	&  $^3\Delta$      &9.36&8.88&8.93&9.12&9.10&9.17&9.17&9.22&9.22&8.87&9.01&8.84 \\
	&  $^3\Sigma^-$    &9.88&9.80&9.80&9.88&9.94&9.92&9.94&9.96&10.02&9.93&9.84&9.74 \\
	%End TOTAL
	\hline
	%Begin TOTAL (dont delete!)
	\multirow{4}{*}{N$_2$}
		& $^3\Sigma^+$     &7.75&7.70&7.65&7.89&7.93&7.79&7.91&7.83&7.72&7.55&7.36&7.20 \\
		&  $^3\Pi$         &8.04&7.77&7.90&8.10&8.01&8.11&8.10&8.17&8.13&8.03&7.79&7.61  \\
		&  $^3\Delta$      &8.88&8.37&8.34&8.61&8.75&8.64&8.79&8.72&8.64&8.39&8.27&8.01 \\
		&  $^3\Sigma^-$    &9.67&7.47&9.40&9.47&9.83&9.62&9.84&9.71&9.67&9.41&9.34&9.13 \\
		%End TOTAL
		\hline
		%Begin TOTAL (dont delete!)
		\multirow{3}{*}{H$_2$CO}
		& $^3\Pi$ 		 &3.50&3.32&3.32&3.42&3.41&3.42&3.48&3.48&3.49&3.41&3.50&3.41 \\
		&  $^3\Sigma^+$    &6.00&5.95&5.93&6.05&6.08&6.02&6.08&6.04&5.94&5.82&6.02&5.91  \\
		&  $^3\Delta$      &7.09&6.32&6.87&7.31&6.40&7.06&6.73&7.09&7.47&7.71&6.98&6.90 \\
		%End TOTAL
		\hline
		%Begin TOTAL (dont delete!)
		\multirow{3}{*}{C$_2$H$_4$}
			&  $^3B_{3u}$    &4.36&4.50&4.45&4.61&4.56&4.48&4.55&4.49&4.30&4.30&4.49&4.43 \\
			&  $^3B_{1u}$    &6.98&6.56&6.89&7.37&6.87&7.19&7.00&7.19&7.44&7.52&7.29&7.13  \\
			&  $^3B_{1u}$    &7.11&6.64&6.98&7.46&6.95&7.30&7.09&7.29&7.57&7.67&7.41&7.26 \\
			%End TOTAL
			\hline
			%Begin TOTAL (dont delete!)
			\multirow{3}{*}{pyridine}
			&  $^3A_1$  &4.05&4.11&4.29&4.41&4.41&4.42&4.41&4.37&4.28&4.17&4.39&4.32   \\
			&  $^3B_1$  &4.41&4.33&4.33&4.51&4.30&4.58&4.44&4.59&4.65&4.60&4.51&4.37  \\
			&  $^3B_2$  &4.56&4.57&4.62&4.70&4.77&4.86&4.83&4.85&4.84&4.75&5.63&5.43   \\
			%End TOTAL
			\hline
			%Begin TOTAL (dont delete!)
			\multirow{3}{*}{benzene}
			&  $^3B_{1u}$  &3.94&4.26&4.22&4.33&4.33&4.26&4.32&4.27&4.19&4.09&4.24&4.17   \\
			&  $^3E_{1u}$  &4.76&4.83&4.90&5.01&4.96&5.04&5.01&5.06&5.07&5.04&5.03&4.94    \\
			&  $^3B_{2u}$  &5.60&5.15&5.24&5.34&5.50&5.55&5.61&5.63&5.65&5.55&5.63&5.43   \\
			%End TOTAL
			\hline	
			\multirow{2}{*}{butadiene}
			&  $^3B_u$     &3.22&3.21&3.23&3.36&3.34&3.32&3.36&3.34&3.26&3.17&3.33&3.23 \\
			&  $^3A_g$     &5.15&5.15&5.07&5.18&5.03&5.11&5.20&5.13&5.06&4.96&5.18&5.13  \\
			%End TOTAL
			\hline
			
   \multirow{4}{*}{}	 
&  MD                  &&$-$0.18&$-$0.11&0.08&0.02&0.07&0.07&0.10&0.10&0.01&$-$0.01&$-$0.14 \\
&  MAD                 &&0.23   &0.18   &0.18&0.17&0.13&0.13&0.13&0.17&0.23&0.21&0.25  \\
&  RMSD                &&0.31   &0.22   &0.22&0.23&0.17&0.18&0.17&0.22&0.29&0.26&0.33  \\
&  \text{Max-Min}      &&1.09   &0.82   &0.65&1.13&0.61&0.74&0.49&0.70&1.11&0.95&1.14  \\
%End TOTAL
		\end{tabular}
	\end{ruledtabular}
	$^a$~All calculations were performed with the aTZVPP AO basis set\cite{TZVP,Kendall_1992}. Upon a reviewer's request, we performed additional aug-cc-pVTZ calculations for TDA-B2PLYP and obtained the same statistics (see Table S8). $^b$ Experimental values  (gas phase) were taken from \citenum{GRIMME_2007}.
	
\end{table*}

In general, we observe in Table~\ref{tab:Grimme_Neese} that most of the results obtained for singlet-triplet excitations have the same good quality as for singlet-singlet excitations. Among our tested methods, $\omega$B2(GP-)PLYP and B2GP-PLYP are the best with RMSDs of only 0.17~eV for the former and 0.18~eV for the latter, followed by RSX-QIDH, $\omega$B97X, and CAM-B3LYP which share the same value of 0.22~eV, a result that is similar to the one for B2PLYP (RMSD = 0.23~eV). PBE-QIDH (RMSD = 0.26~eV) and RSX-0DH (RMSD = 0.29~eV) come next, followed by poorly performing B3LYP (RMSD = 0.31~eV) and PBE0-DH (RMSD = 0.33~eV). If we take a closer look at the MDs, we note that B3LYP has a tendency to underestimate the singlet-triplet excitation energies along with CAM-B3LYP, PBE-QIDH, and PBE0-DH. Contrary to that, most of the other DFAs tend to overestimate these energies within one-tenth of an eV. While B2PLYP has an MD of 0.02 eV, its MAD of 0.17 eV indicates that excitations are equally under- and overestimated. The same can also be said for PBE-QIDH (MD = $-$0.01 eV; MAD = 0.21 eV) and RSX-0DH (MD = 0.01 eV; MAD = 0.23 eV).

Another interesting aspect is to assess the influence of the LC scheme on the singlet-triplet excitation energies. From a total of 22 transitions, \text{$\omega$B2(GP-)PLYP} both present an improvement for 14, reducing the deviation for some systems by up to 50\%. On the other hand, ethene and pyridine are cases that present an opposite trend. For instance, in ethene the only DFA that gives a small energy deviation for the two $^3B_{1u}$ states is B2GP-PLYP ($\pm$0.02~eV) followed by CAM-B3LYP (around $-$0.12~eV) and B2PLYP (around $-$0.13~eV). On the other hand, $\omega$B2(GP-)PLYP and RSX-(QI/0)DH all exhibit blueshifts for the same two excitations that range from 0.1 to 0.2~eV.

We conclude from this section that the LC scheme seems to be beneficial for most cases and for all tested DHDFs. However, one potential drawback is the reliance on experimental values in this set. We have only included these data here to establish a connection with the early TDA-B2PLYP investigation from 2007, and in the following we focus on high-level wave-function reference values and compare TD-DFT with TDA-DFT approaches.  

\subsection{The updated Gordon benchmark set}
\label{sec:gordon}

In the following sections, we assess the vertical triplet excitation energies in two different ways: first, we begin with a thorough analysis of  all singlet-triplet excitation energies in the given set, i.e. the transition from the S0 to various triplet states of the same molecule, labeled here as "T$N$" with $N\geq$1. This is then followed by a separate analysis of their first singlet-triplet energy splitting (S1-T1). The latter may be particularly useful for testing a method's robustness prior to any applications of current technological relevance, as we already outlined in Section \ref{sec:compdet}. Such an analysis also excludes any potential high-lying states for which DHDFs are known to be occasionally problematic due to the reliance on the CIS(D) correction.\cite{Goerigk_2009,TDDHDFaccount} Our S0-T$N$ and S1-T1 analyses in the remainder of this paper will provide a description of the triplet excitations of contemporary DFAs based on accurate high-level data, contrary to the previous section which relied on various experimental data.

We begin our discussion with the updated Gordon set. Reference values were determined in 2017 by Schwabe and Goerigk for both singlet-singlet and singlet-triplet excitations, but only the first were published in Ref.~\citenum{Schwabe_2017}. In this work, we present the first analysis of Schwabe and Goerigk's estimated and full CC3/aug-cc-pVTZ data with a total of 38 electronic excitations made up of 27 valence and 11 Rydberg excitations from 12 different molecules; see Supplementary Material for the reference values and Ref.~\citenum{Schwabe_2017} for details on how they were obtained. Our analysis focusses on the same TD-DFT and TDA-DFT methods discussed before, but in addition we also report values for linear-response coupled-cluster singles and doubles (CCSD)~\cite{ccsd} in the framework of linear response theory. Those numbers were a by-product of the CC3 calculations carried out to obtain the reference values.\cite{Schwabe_2017}  CCSD is computationally more expensive than DHDFs with a formal scaling behavior of $\mathcal{O}(N^6)$ compared to $\mathcal{O}(N^5)$, with $N$ being the system size or number of AOs. If DHDFs turn out to have statistics very similar to CCSD, it would constitute an important milestone. 

\subsubsection{All singlet-triplet excitations in this set (S0-T$N$)}

\begin{figure}
	\centering
	\includegraphics[width=1.0\linewidth]{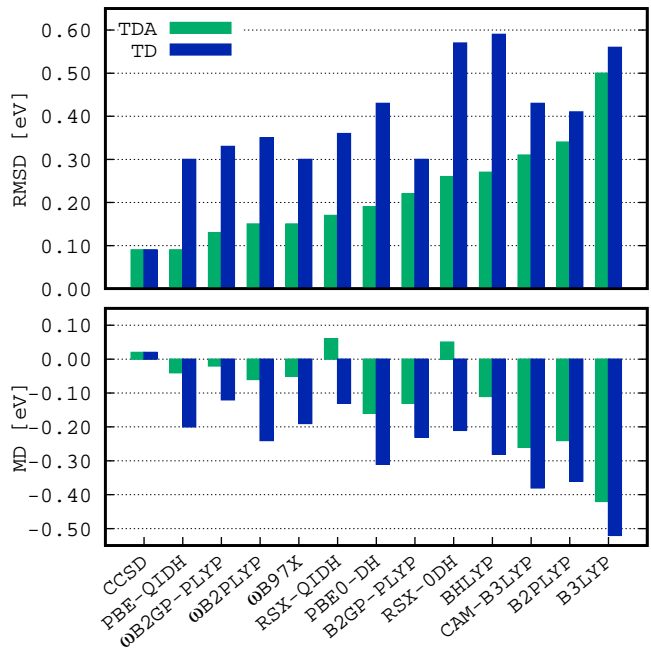}
	\caption{RMSDs (top) and MDs (bottom) over all vertical singlet-triplet excitations in the updated Gordon benchmark set using TDA (green) and TD(blue).The aug-cc-pVTZ AO basis set was applied in all cases. All values are in eV.}
	\label{fig:gordon_1}
\end{figure}

From Figure~\ref{fig:gordon_1} we can immediately conclude that there are considerable differences between the TDA and TD algorithms. Overall, TD fails to describe the singlet-triplet excitations due to a systematic red-shift and regardless of whether the DFA has been LR-corrected or not. This tendency mirrors findings reported previously for conventional global and LC-hybrids.\cite{triplet_inestability2011,triplet_inestability2017,Jacquemin2017_2,Tozer_2013_TDA} Interestingly, the difference between the TDA and TD frameworks seem to be larger for LC-DFAs than for global DFAs. For instance, the largest difference is observed for $\omega$B2PLYP, where the RMSD has increased by 0.13~eV compared to its global cousin. Similarly, $\omega$B2GP-PLYP and \mbox{RSX-0DH} display the same behavior with differences of 0.09 and 0.07~eV, respectively.  Despite this drawback, the inclusion of an LC scheme still seems to be beneficial for some of the tested LC-DFAs as discussed below. Based on our findings, we focus any subsequent discussions and analyses on TDA results; however, we continue reporting TD results in the figures and the Supplementary Material.

We begin our discussion by analyzing the RMSDs resulting from TDA-DFT calculations. As all excitations except one (water) are of local-valence character, the best DFA is PBE-QIDH (RMSD = 0.09~eV) which performs as well as CCSD. This result is followed by $\omega$B2GP-PLYP (RMSD = 0.13~eV). $\omega$B2PLYP and $\omega$B97X share the same value of 0.15~eV. RSX-QIDH and PBE0-DH follow with RMSDs of 0.17~eV and 0.19~eV, respectively. Note that the six aforementioned DFAs are the only ones that show RMSDs below 0.2~eV. B2GP-PLYP, RSX-0DH, and BHLYP follow with RMSDs between 0.2 and 0.3 eV. The three worst DFAs with RMSDs above 0.3~eV are CAM-B3LYP, B2PLYP, and B3LYP. We therefore see the previously observed trend that the positive influence of the LC is limited to the BLYP-based DHDFs.

In the case of MDs, every single DFA---with the exception of RSX-(QI/0)DH---tends to underestimate the excitation energies. CCSD slightly overestimates the excitation energies with a near-perfect MD of 0.02~eV, whereas \text{$\omega$B2GP-PLYP} performs very similarly with a negative MD of $-$0.02~eV. Those results are closely followed by PBE-QIDH (MD = $-$0.04), RSX-0DH (MD = 0.05~eV), $\omega$B97X (MD = $-$0.05~eV), RSX-QIDH (MD = 0.06) and  $\omega$B2PLYP (MD = $-$0.06~eV). Every other DFA exhibits systematic underestimation ranging from $-$0.11 eV for BHLYP to $-$0.42~eV for B3LYP.

\subsubsection{First singlet-triplet energy splitting (S1-T1)}

\begin{figure}
	\centering
	\includegraphics[width=1.0\linewidth]{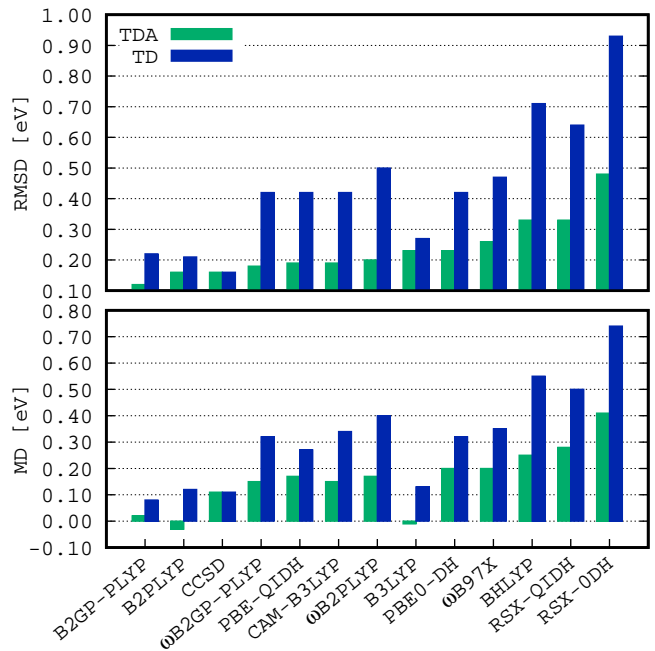}
	\caption{RMSDs (top) and MDs (bottom) for the S1-T1 splittings in the updated Gordon benchmark set using TDA (green) and TD(blue).The aug-cc-pVTZ AO basis set was applied in all cases. All values are in eV.}
	\label{fig:gordon_2}
\end{figure}

According to Figure~\ref{fig:gordon_2}, and considering only TDA values for the reasons stated in the previous section, B2GP-PLYP and B2PLYP are the two DFAs with the best RMSDs of just 0.12 and 0.16~eV, respectively. They are closely followed by CCSD (RMSD = 0.16~eV), $\omega$B2GP-PLYP (RMSD = 0.18~eV), PBE-QIDH(RMSD = 0.19~eV), CAM-B3LYP (RMSD = 0.19~eV), and $\omega$B2PLYP (RMSD = 0.20~eV). Surprisingly, B3LYP and PBE0-DH (RMSD = 0.23~eV in both cases) outperform $\omega$B97X (RMSD = 0.26~eV), which usually ranked well in our previous sections. All remaining methods, i.e. BHLYP, RSX-QIDH, and RSX-0DH, yield RMSDs above 0.3~eV and do not seem suitable. 

We would like to point out that TDA-B2(GP-)PLYP outperform their LC-counterparts when the transition involves local-valence excitations. Since most of the systems in this analysis have local-valence S1 and T1 states, it is understandable that global DHDFs also perform well as shown in our earlier sections. For example, the 1$^1$A$_u$ state in glyoxal has an energy deviation from the reference of only 0.02~eV for B2GP-PLYP, which increases to 0.09~eV for $\omega$B2GP-PLYP. Benzene is another problematic case and, independent of the chosen DFA, the already exisiting overestimation of the 1$^1$B$_{2u}$ state worsens because of the inclusion of the LC (Tables S11 and S12). On the other hand, the 1$^1$B$_1$ state of water is a Rydberg excitation for which $\omega$B2(GP-)PLYP exhibit a reduction in the energy deviation. For instance, B2GP-PLYP shows a severe underestimation of $-$0.37~eV which decreases to $-$0.20~eV by the inclusion of the LC scheme.

The MDs in Figure~\ref{fig:gordon_2} are close to 0 eV for B3LYP (MD = $-$0.01~eV), B2GP-PLYP (MD = 0.02~eV), and B2PLYP (MD = $-$0.03~eV). Every other tested method displays systematic overestimation, incl. CCSD (MD = 0.11~eV). While most MDs range from 0.15~eV (CAM-B3LYP and $\omega$B2GP-PLYP) to 0.2~eV (PBE-0DH and $\omega$B97X),   BHLYP and RSX-(QI/0)DH suffer from severe overestimation with MDs larger than 0.25~eV and RSX-0DH having the largest MD with a value of 0.41~eV.

We therefore see that the assessed BLYP-based DHDFAs are comparable or sometimes even better than the more costly CCSD. While global BLYP-based DFAs may be sufficient for S1-T1 transitions of the local-valence type, the LC-counterparts are needed for any transitions with partial LR character, such as Rydberg excitations.

\subsection{The Thiel benchmark set}
\label{sec:thiel}

To investigate the TD(A)-DHDF's applicability, we also present results for the popular benchmark set by Thiel and co-workers.\cite{Thiel2008} The original set comprises 167 valence excitations, divided into 104 singlet and 63 triplet excitations, with various references values out of which we chose the CC3/TZVP~\cite{TZVP} ones, for reasons outlined in Ref. \citenum{Schwabe_2017}. The singlet-singlet excitations have been thoroughly addressed with TD(A)-DHDFs in previous articles---see Refs. \citenum{Goerigk_Moellmann_2009}, \citenum{Schwabe_2017}, and \citenum{ADAMOJCC2020}---while the singlet-triplet excitations have never been investigated with them. However, most of the molecules in the set have already been addressed in our preceding analysis and we decided to only analyze those six molecules that have not been assessed before, namely acetamide, benzoquinone, imidazole, naphthalene, norbornadiene, and octatetraene. For those six molecules, 24 local-valence excitations needed to be analyzed. Additional linear-response CCSD data have been taken from Ref.\citenum{Thiel2008}. All results are shown in Tables S13 and S14 in the Supplementary Material. 

\subsubsection{All singlet-triplet excitations in this set (S0-T$N$)}

\begin{figure}
	\centering
	\includegraphics[width=1.0\linewidth]{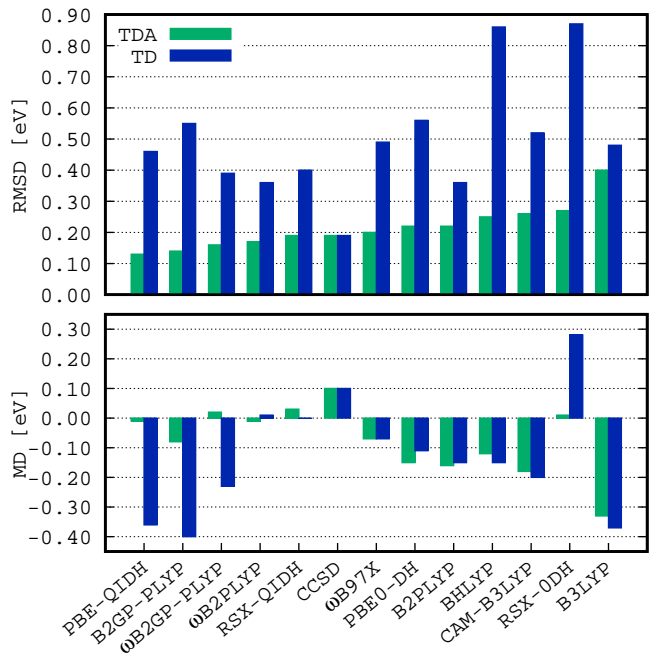}
         \caption{RMSDs (top) and MDs (bottom) over vertical singlet-triplet excitations in six molecules of the Thiel benchmark set using TDA (green) and TD(blue). The TZVP AO basis set was applied in all cases. All values are in eV.}
\label{fig:thiel_1}
\end{figure}

Again, we can confirm that TD-DFT methods are unable to adequately describe singlet-triplet excitations (see Figure~\ref{fig:thiel_1}). Most of the DFAs consistently underestimate the excitation energies, as shown by the MDs, with B2GP-PLYP (MD = $-$0.40~eV), B3LYP (MD = $-$0.37~eV), and PBE-QIDH (MD = $-$0.36~eV) representing the most severe cases. RSX-0DH is the only DFA that displays a (large) overestimation of 0.28~eV. In contrast, RSX-QIDH has a perfect MD of 0.00~eV, followed by $\omega$B2PLYP with a slight overestimation of 0.01~eV. The MDs for the other TD-DFT approaches range from $-$0.11 to $-$0.23~eV. 

Switching to the TDA-DFT algorithm induces a strong blueshift in some DFAs, resulting in very good MDs for PBE-QIDH (MD = $-$0.04~eV), B2GP-PLYP (MD = $-$0.08~eV), $\omega$B2GP-PLYP (MD = 0.02~eV), RSX-0DH (MD = 0.01~eV), $\omega$B97X (MD = $-$0.07~eV), and $\omega$B2PLYP (MD = $-$0.01~eV). Interestingly, RSX-QIDH is not much affected by changing its algorithm and still displays a reasonably low MD of 0.03. All those DFAs have in common that they outperform CCSD (MD = 0.10~eV). From now on we will only discuss the TDA-DFT results. 

The overall trend for the TDA-based RMSDs is shown in Figure~\ref{fig:thiel_1}, and PBE-QIDH ranks as the best DFA with a value of 0.13~eV, closely followed by B2GP-PLYP (RMSD = 0.14~eV) and its LC-counterpart $\omega$B2GP-PLYP (RMSD = 0.16~eV). Meanwhile, $\omega$B2PLYP and RSX-QIDH also perform well with RMSDs of 0.17~eV and  0.19~eV, respectively. All the aforementioned methods compete with or outperform CCSD (RMSD = 0.19~eV).  With the exception of PBE0-DH and RSX-0DH (RMSD = 0.22 and 0.27~eV), we see that the best DFAs for this set are exclusively DHDFs, with the hybrid $\omega$B97X having an RMSD of 0.20~eV, while the hybrids BHLYP and CAM-B3LYP display very similar values of 0.25 and 0.26~eV, and B3LYP ranks once again in last place with an RMSD of 0.40~eV.

\subsubsection{First singlet-triplet energy splitting (S1-T1)}

\begin{figure}
	\centering
	\includegraphics[width=1.0\linewidth]{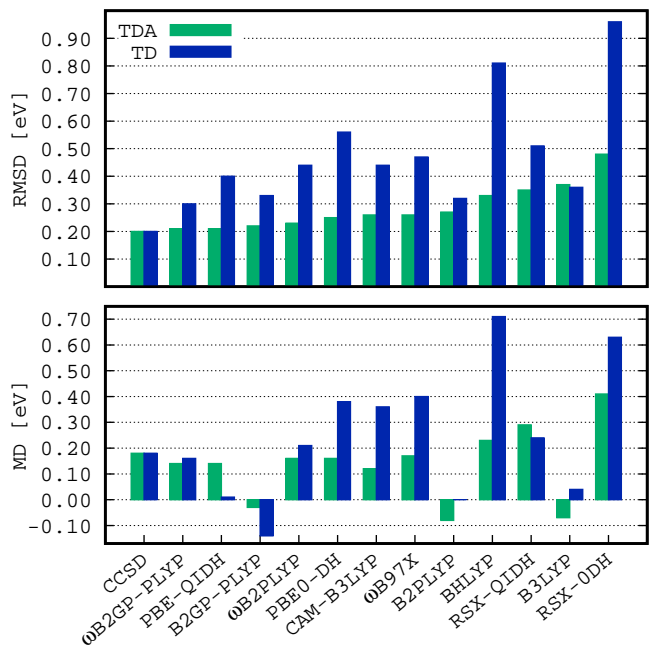}
	\caption{RMSDs (top) and MDs (bottom) for the S1-T1 splittings in six molecules of the Thiel benchmark set using TDA (green) and TD(blue).The TZVP AO basis set was applied in all cases. All values are in eV.}
	\label{fig:thiel_2}
\end{figure}

The first singlet-triplet energy splitting (S1-T1) in Figure~\ref{fig:thiel_2} shows that this time no DFA has an RMSD below 0.2~eV, in contrast to the previous section. Let us point out that CCSD is the only method that has a value of exactly 0.20~eV. However, $\omega$B2GP-PLYP and PBE-QIDH are the best DFAs sharing the same RMSD of 0.21~eV, followed by B2GP-PLYP and $\omega$B2PLYP, which display RMSDs of 0.22 and 0.23~eV, respectively. PBE0-DH, CAM-B3LYP, $\omega$B97X and B2PLYP follow with RMSDs ranging from  0.25 to 0.27~eV.  The remaining methods have RMSDs that range from   0.33 to 0.48~eV. Surprisingly, RSX-QIDH (RMSD = 0.35~eV) is among them despite  being one of the better methods for the S0-T$N$ transitions. This indicates how the analysis of S1-T1 gaps informs on a method's robustness. If a S0-T1 transition is described reasonably well, but the S0-S1 is not---we saw in Section \ref{sec:singlets} how RSX-QIDH was outperformed by many other methods for singlet-singlet transitions---the S1-T1 gap is consequently not well described. We can see this for RSX-QIDH where the RMSD for all S0-S1 gaps for the six molecules of the Thiel set is 0.32~eV, but for all S0-T1 0.13~eV. Contrary to that, B2GP-PLYP has RMSDs of 0.24~eV for S0-S1 and 0.11~eV for S0-T1, respectively. 

Analyzing the MDs, we observe that B2GP-PLYP has the lowest value of $-$0.03~eV, while having an MAD of 0.20~eV, which indicates that the MD is a result of some error compensation. On the other hand, $\omega$B2GP-PLYP seems to be more robust with an MD of 0.14~eV and an MAD of 0.19~eV, indicating a systematic overestimation of the S1-T1 gaps. The same holds for PBE-QIDH, which also has an MD of 0.14~eV and an MAD of 0.19~eV. The second DFA in the MD ranking is B3LYP with an underestimation of 0.07~eV but, similarly to B2GP-PLYP, a very different MAD (0.34~eV). $\omega$B97X seems to be more robust, systematically overestimating the gaps (MD=0.17~eV and MAD=0.22~eV). Therefore, it is important to keep in mind that DFAs can emerge with good MDs due to compensation between under and overestimated excitation energies, but this ought be accompanied by an analysis of their MADs too. Note that the discussed DFAs present lower MD values compared to CCSD, which has a value of 0.18~eV for both MD and MAD, indicating a systematic overestimation (see full details in the Supplementary Material).

\subsection{The Loos and Jacquemin benchmark set}
\label{sec:loos}

The original benchmark set proposed by Loos, Jacquemin, and co-workers~\cite{loos2018} consists of 47 singlet-triplet transitions of varying character, i.e. local valence or Rydberg. However, for our work we have selected only 34 transitions to 23 local valence, 11 Rydberg, and one CT states in 13 different molecules, which had not been included in any of the previously discussed sets. We mention in passing that the triplet state for the only CT system, namely HCl, does not form part of the original benchmark set and we calculated it in the present work following the same strategy as in Ref. \citenum{loos2018}. All reference values are of CC3/aug-cc-pVTZ quality. All linear-response CCSD values were taken from Ref. \citenum{loos2018} with the exception of the one CT transition that we calculated. All values for this set are shown in Tables S17 and S18. In this section, we only discuss TDA-DFT results, while TD values are shown in the figures.

\subsubsection{All singlet-triplet excitations in this set (S0-T$N$)}

\begin{figure}
	\centering
	\includegraphics[width=1.0\linewidth]{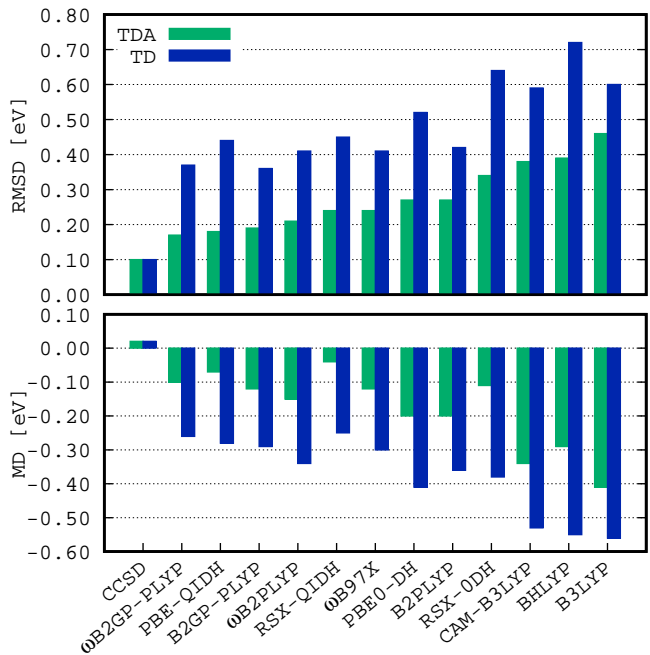}
         \caption{RMSDs (top) and MDs (bottom) over vertical singlet-triplet excitations in 13 molecules of the Loos and Jacquemin benchmark set using TDA (green) and TD(blue). The aug-cc-pVTZ AO basis set was applied in all cases. All values are in eV.}
	\label{fig:loos_1}
\end{figure}

Figure~\ref{fig:loos_1} shows the same trend as for the two previous benchmark sets, showing the robustness and reliability of our study. This time, CCSD exhibits an RMSD of 0.1~eV and we only identify three DFAs with RMSDs below 0.2~eV, namely $\omega$B2GP-PLYP with the lowest value of 0.17~eV, followed by PBE-QIDH and B2GP-PLYP with values of 0.18 and 0.19~eV, respectively. Most of the remaining DFAs yield values ranging between 0.2 and 0.3~eV: $\omega$B2PLYP (RMSD = 0.21~eV), RSX-QIDH and $\omega$B97X (RMSD = 0.24~eV), and PBE0-DH and B2PLYP (RMSD = 0.27~eV). RSX-0DH, CAM-B3LYP, BHLYP and B3LYP perform the worst in this section, yielding the largest RMSDs
ranging from 0.34 to 0.46~eV.

Most DFAs display systematic underestimation, similarly to what we have seen before. Despite the fact that RSX-QIDH is the DFA with the lowest absolute MD ($-$0.04~eV), its MAD almost reaches 0.2~eV. PBE-QIDH and $\omega$B2GP-PLYP appear to be more balanced, since they have values of $-$0.07 and $-$0.10~eV for the MDs, and 0.15 and 0.14~eV for the MADs, respectively.

\subsubsection{First singlet-triplet energy splitting (S1-T1)}

\begin{figure}
	\centering
	\includegraphics[width=1.0\linewidth]{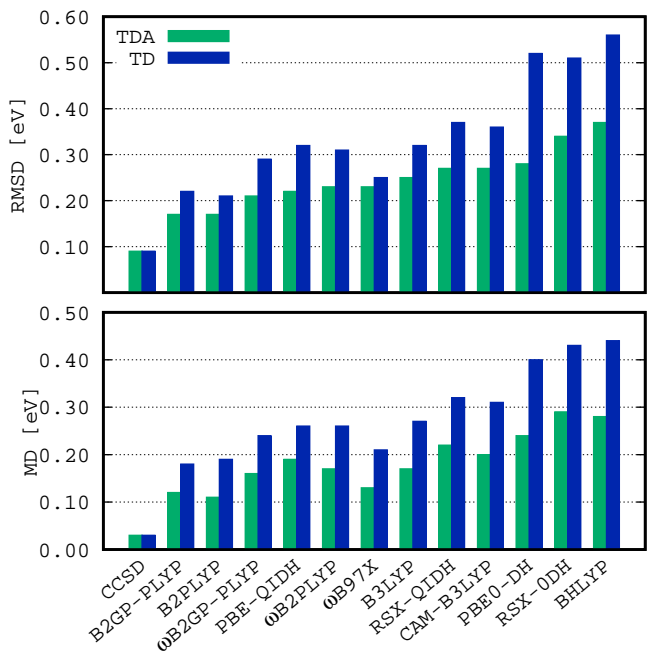}
	\caption{RMSDs (top) and MDs (bottom) for the S1-T1 splittings in 13 molecules of the Loos and Jacquemin benchmark set using TDA (green) and TD(blue). The aug-cc-pVTZ AO basis set was applied in all cases. All values are in eV.}
	\label{fig:loos_2}
\end{figure}

For the S1-T1 energy splitting (Figure~\ref{fig:loos_2}) we see that CCSD has the lowest RMSD with a value of 0.09~eV and that B2(GP-)PLYP have the same---and the best---RMSDs amongst the DFAs both with a value of 0.17~eV; they are, in fact, the only two DFAs with RMSDs below 0.2~eV. $\omega$B2GP-PLYP, PBE-QIDH, $\omega$B2PLYP, as well as $\omega$B97X follow next with values of 0.21 for the first, 0.22 for the second, and 0.23~eV for the last two DFAs, respectively. Surprisingly, B3LYP (RSMD = 0.25~eV)  surpasses RSX-QIDH and CAM-B3LYP, which share the same value of 0.27~eV, as well as PBE0-DH with a value of 0.28~eV. Ultimately, and like in most of our previously discussed cases, RSX-0DH and BHLYP occupy the last two places with RMSDs of 0.34 and 0.37~eV, respectively.

Interestingly, and similarly to our previous discussion on singlet-singlet excitations in Sections \ref{sec:training} and \ref{sec:gordon_sing}, we see that some Rydberg states do not need require the LC, contrary to our previously discussed sets. Instead, global DFAs such as B2(GP-)PLYP are already able to describe them, while LC DFAs tend to overestimate them. For instance, in carbon monoxide the  $^3 \Sigma^+$ state is already overestimated by 0.15~eV with B2GP-PLYP, which is further blueshifted by an additional 0.13 eV for $\omega$B2GP-PLYP. Another example for a system displaying such behavior is diazomethane, where the $^3 B_1$ state is overestimated by 0.04~eV with B2GP-PLYP and by 0.15~eV with $\omega$B2GP-PLYP. PBE-QIDH overestimates the same state by 0.28~eV, which is further worsened by an additional 0.14~eV through the LC in RSX-QIDH.

As already mentioned before, the MD shows a systematic blueshift in the energy splittings. B2PLYP and the related B2GP-PLYP give the smallest MDs with 0.11 and 0.12~eV, respectively, closely followed by $\omega$B97X with an MD of 0.13~eV. The LC-DFAs $\omega$B2GP-PLYP and $\omega$B2PLYP come next with values of 0.16 and 0.17~eV, and the latter in turn shares the same value with B3LYP. Finally, PBE-QIDH has an MD of 0.19~eV, being the last DFA with a value below 0.2~eV. The remaining DFAs, namely CAM-B3LYP, RSX-QIDH, PBE0-DH, BHLYP, and RSX-0DH display results with larger MDs of 0.20, 0.22, 0.24, 0.28, and 0.29~eV, respectively. In contrast to the previous set, CCSD seems to be more balanced this time displaying a slight on-average overestimation with an MD of 0.03~eV and an MAD of 0.07~eV. In general, the S1 state suffers from a blueshift while the T1 suffers from a redshift. This results in the observed blueshifts for the S1-T1 gaps.

\subsection{The ``exotic-molecules'' set}
\label{sec:exotic_trip}

As previously discussed in Section \ref{sec:exotic_sing}, this benchmark set contains species that can be considered "exotic" for organic chemistry (F, Cl, Si, or P atoms). Therefore, it seems reasonable to also analyze their performance for triplet excited states.  All values are also shown in Tables S21and S22. Linear-response CCSD numbers were taken from Ref. \citenum{exotic2020}. Only TDA-DFT results are discussed for this set, while TD-DFT results are shown in the figures.

\subsubsection{All singlet-triplet excitations in this set (S0-T$N$)}

\begin{figure}
	\centering
	\includegraphics[width=1\linewidth]{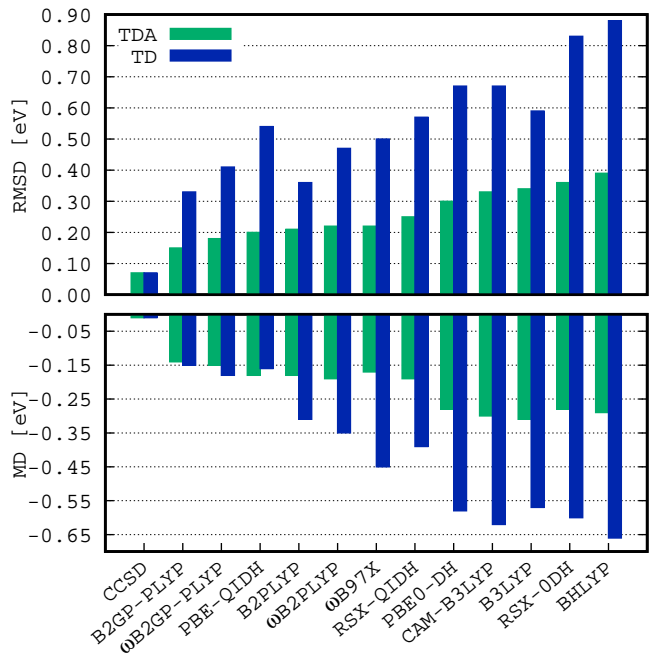}
  \caption{RMSDs (top) and MDs (bottom) over vertical singlet-triplet excitations in the ``exotic-molecules'' benchmark set using TDA (green) and TD(blue). The aug-cc-pVTZ AO basis set was applied in all cases. All values are in eV.}
	\label{fig:loos2_trip}
\end{figure}

The results are shown in Figure~\ref{fig:loos2_trip} and the general trend seems to be the same as for the other sets. CCSD is the method with the lowest RMSD (0.07~eV), whereas B2GP-PLYP and its related LC-version have the lowest RMSDs of all DFAs with values of 0.15 and 0.18~eV, respectively. Next, we see that PBE-QIDH, B2PLYP and $\omega$B2PLYP perform very similarly with RMSDs of 0.20, 0.21 and 0.22~eV, respectively. Again, $\omega$B2PLYP yields the same value as $\omega$B97X. The ``non-empirical'' RSX-QIDH comes next with an RMSD of 0.25~eV, being the last DFA with a value lower than 0.3~eV. On the other hand, PBE0-DH has an RMSD of 0.30~eV,  and CAM-B3LYP a value of 0.33~eV, just one hundredth of an eV below B3LYP (RMSD = 0.34~eV). Finally, and by now unsurprisingly, the last two DFAs in the list are RSX-0DH along with BHLYP, with values of 0.36 and 0.39~eV, respectively.

Similarly, there is a systematic underestimation in the transitions. CCSD presents the lowest absolute value with an MD of just $-$0.01~eV whereas most of the DFAs present absolute values below 0.2~eV, with B2GP-PLYP being the best (MD = $-$0.14~eV) and $\omega$B2PLYP along with RSX-QIDH the last ones in this energy range (MD = $-$0.19). All other MDs range from $-$0.28~eV (RSX-0DH and PBE0-DH) to $-$0.31 eV (B3LYP). For this set, we therefore see only very small impact of the LC on all tested DHDFs, which mirrors our findings for the singlet-singlet excitations in Section \ref{sec:exotic_sing}.

\subsubsection{First singlet-triplet energy splitting (S1-T1)}

\begin{figure}
	\centering
	\includegraphics[width=1\linewidth]{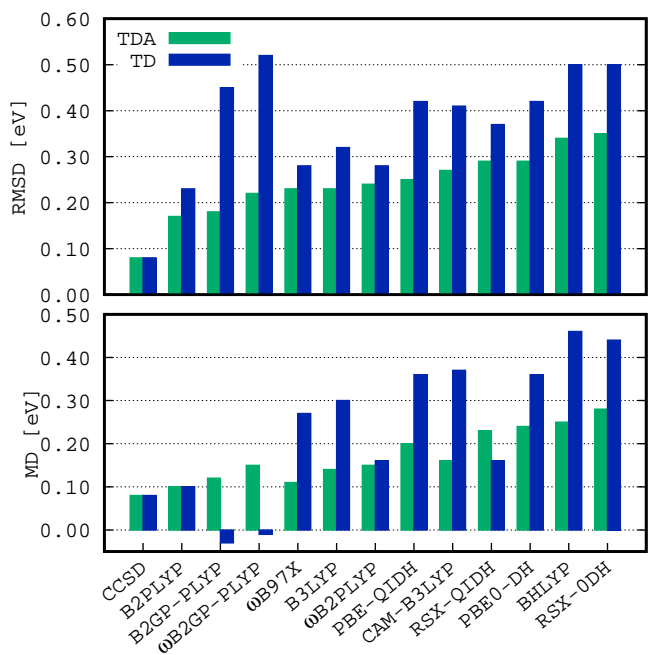}
	\caption{RMSDs (top) and MDs (bottom) for the S1-T1 splittings  in the ``exotic-molecules'' benchmark set using TDA (green) and TD(blue). The aug-cc-pVTZ AO basis set was applied in all cases. All values are in eV.}
	\label{fig:loos2_trip2}
\end{figure}

Our final discussion in this work is the S1-T1 energy splitting for this set. In Figure~\ref{fig:loos2_trip2} we observe that B2(GP-)PLYP perform best with values of 0.17(0.18~eV), being the only two methods with energy deviations lower than 0.2~eV. $\omega$B2GP-PLYP follows with an RMSD of 0.22~eV, followed by $\omega$B97X and B3LYP, which share the same value of 0.23~eV, and $\omega$B2PLYP with a value of 0.24~eV. PBE-QIDH yields an RMSD of 0.25~eV. This compares with a value of 0.29~eV for RSX-QIDH, showing that the LC increases the S1-T1 on average for both the BLYP and PBE-based methods. We also see this when comparing PBE0-DH (RMSD = 0.29~eV) with RSX-0DH (RMSD = 0.35~eV).

We conclude this discussion by coming back to our TD-DFT results, which deliver a different picture for the S1-T1 gaps. For instance, the worst functional for the TD-DFT scheme is now $\omega$B2GP-PLYP, which has always displayed a very good performance in every previous sets regardless of the chosen algorithm, with an RMSD of 0.52~eV. Contrary to that, B2PLYP still remains in first place with a marginally increased value of 0.23~eV; a slightly higher value is seen for its LC version $\omega$B2PLYP (0.28~eV). More interestingly, B2GP-PLYP, one of the best DFAs for S1-T1 energy splittings with TDA-DFT, also worsens for the TD-DFT algorithm. A comparison of the MDs for the whole set in Figure \ref{fig:loos2_trip} with those for the S1-T1 gap in Figure \ref{fig:loos2_trip2} reminds us, that triplet states are usually underestimated by the TD-DFT scheme. This ultimately leads to an on-average enlarged S1-T1 gap.

\section{Summary and outlook}
\label{sec:conclusions}

With one exception, all previous excited-state studies with double-hybrid density functionals (DHDFs) using Grimme and Neese's definition\cite{GRIMME_2007} from 2007 have exclusively dealt with singlet-singlet transitions and left out technologically important triplet excited states. Moreover, all previous studies, except for two,\cite{GRIMME_2007,Schwabe_2017} have either relied entirely on a time-dependent (TD) or a Tamm-Dancoff-Approximation (TDA) Density Functional Theory (DFT) algorithm. Most TD(A)-DHDFs have been based on global hybrid-components that relied either on BLYP and PBE expressions for the DFT exchange and correlation components. While some of them---such as B2PLYP, B2GP-PLYP, PBE-QIDH or various spin-component and spin-opposite scaled methods (see Ref. \citenum{TDDHDFaccount} for examples)---have shown excellent results, they usually failed in the description of charge-transfer (CT) transitions. The reason for this failure is the same as for global hybrid-DFT methods, namely the wrong decay of the exchange-correlation potential with increasing electron-electron distance, which cannot be simply fixed by the second-order perturbative correction in TD(A)-DHDFs.\cite{wB2PLYPwB2GPPLYP} In 2019, we addressed this problem by presenting the first two TD-DHDFs following Grimme and Neese's definition that included a long-range correction (LC) in their exchange part and were optimized for excitation energies.\cite{wB2PLYPwB2GPPLYP} We named those methods $\omega$B2PLYP and $\omega$B2GP-PLYP, as they were extensions of the exisiting and well-performing B2PLYP and B2GP-PLYP global DHDFs. $\omega$B2PLYP and $\omega$B2GP-PLYP produced some of the best-reported TD-DFT results for organic molecules and were a significant improvement over global density functional approximations (DFAs), incl. global DHDFs, for Rydberg and charge-transfer (CT) transitions. However, at that time we had been unable to test those methods for singlet-triplet excitations, nor have we assessed them within the TDA or compared them with ground-state optimized LC-DHDFs that relied on PBE components.

In this work, we followed up on our previous contribution and closed the aforementioned gaps. We compared global and our LC-DHFS based on BLYP exchange-correlation expressions with PBE-based ones within both the TDA and TD-DFT frameworks. To our knowledge, this has been the first time that the ``non-empirical'' LC-DHDFs RSX-QIDH and RSX-0DH have been applied to excited-state problems within both the TDA and TD-DFT contexts. We tested a total number of eight DHDFs, but with the exception of B2PLYP, none of the others had ever been assessed for singlet-triplet excitations. Comparisons were also made with popular global and LC-hybrid DFAs.

Our study began with revisiting singlet-singlet excitations to allow for the analysis of the PBE-based methods and the TDA scheme. For that purpose, we compared our new TDA- and TD-DFT data with our previously published\cite{wB2PLYPwB2GPPLYP} TD-($\omega$)B2(GP-)PLYP results on two test sets that considered local-valence and Rydberg excitations---one of them being the updated Gordon benchmark set by Schwabe and Goerigk\cite{Schwabe_2017}---and a modified set for CT excitations. For the first time we also analyzed the fairly new ``exotic-molecules''' benchmark set by Loos, Jacquemin and co-workers\cite{exotic2020} with DFT methods; this local-valence test set contains main-group elements that are more unusual compared to standard organic-molecule benchmark sets. The influence of the TDA seems to be stronger for local-valence excitations than for Rydberg excitations, but in general both the TDA and TD algorithms can be recommended. Safe choices for such excitations seem to be the LC-DHDFs $\omega$B2PLYP and $\omega$B2GP-PLYP as well as the global DHDFs PBE-QIDH, B2GP-PLYP, and PBE0-DH. Note however, how B2GP-PLYP is safer to use for local-valence than for Rydberg excitations according to our results, while PBE0-DH may be less applicable to larger chromophores according to Ref. \citenum{Goerigk_2014}. For the BLYP-based DHDFs, the LC had a positive impact on TD-DHDFs for both local-valence and Rydberg transitions, as also shown in Ref. \citenum{wB2PLYPwB2GPPLYP}. Moreover, we reported a positive impact of the LC on Rydberg excitations but a minor negative impact for local-valence excitations within the TDA formalism. The latter was mostly due to the fact that  \mbox{$\omega$B2(GP-)PLYP} had been fitted within the TD-DFT framework. However, in the bigger picture, they were still applicable and perform better than many of the other tested methods. Interestingly, the LC had the opposite effect on the PBE-based methods regardless of the type of transition or algorithm. Despite the relatively good performance of PBE-QIDH and PBE0-DH, those methods did not describe CT excitations well, nor did any of the other global DFAs. The LC  had a negative impact on the PBE-based methods, did not seem to solve the CT problem for them, and instead resulted in severe overestimation of the excitation energies. More robust statistics were found for the BLYP-based LC-DHDFs. Therefore, we recommend $\omega$B2PLYP and $\omega$B2GP-PLYP whenever any potential CT contribution to an excitation is expected.

Our analysis of excitations from singlet ground states to various triplet excited states comprised various test cases, starting with a repetition of Grimme and Neese's early work for a small set relying on experimental reference data. We then shifted our focus to reliable, high-level wave-function reference values and presented hitherto unpublished values for the updated Gordon benchmark set generated during a study by our and the Schwabe groups in 2017.\cite{Schwabe_2017} The findings for this set were then complemented by additional molecules from test sets by Thiel as well as Loos and Jacquemin. Our results confirmed previously established findings\cite{triplet_inestability2011,triplet_inestability2017,Jacquemin2017_2,Tozer_2013_TDA} that the usual TD-DFT algorithm was not suitable for such excitations and that the TDA had to be employed in its stead. The presence of the CIS(D) correction in TD-DHDFs does not change this problem. Most of our assessed singlet-triplet excitations had local-valence character and our findings resembled the ones discussed for the singlet-singlet excitations with the exception of PBE0-DH, which we cannot recommend any longer. Again, the LC had a smaller and mostly beneficial impact on BLYP-based functionals contrary to the PBE-based ones. The well-performing DFAs were at times also competitive with the more expensive CCSD approach.

The analysis of S1-T1 gaps was particularly informative and showed that there is still room for future improvement to achieve true robustness. Often, a method that overestimated S0-S1 transitions might at the same time underestimate S0-T1 transitions, which led to an overestimation of the S1-T1 gaps. This was particularly the case for LC-DHDFs that tended to slightly overestimate singlet-singlet local-valence transitions, but was different when Rydberg states were involved. It is therefore difficult to recommend just one method that works well for all scenarios. However, until such a method has been developed, we recommend BLYP-based global DHDFs and their LC counterparts based on an analysis of MDs, MADs, RMSDs and error ranges.

Our work has provided an overview of state-of-the-art TD(A)-DHDFs for applications to both singlet-singlet and singlet-triplet excitation energies. Achieving a simultaneously accurate description of both is challenging, but the TDA variants of BLYP-based global and LC-DHDFs can be recommended, as they tend to outperform the currently applied hybrid-DFT methods. The calculation of singlet-triplet excitation energies with TD(A)-DHDFs will be possible for the general user with the next ORCA release, which may encourage changes to current practice in typical applications. Our work also inspired further developments to both solve the triplet-state problem of TD-DFT and to enable a consistent description of excitation energies below the chemical-accuracy threshold with TD(A)-DHDFs. We hope to report further improvements in the near future and to extend our studies to even larger test sets and other systems.

\section*{Supplementary Material}
The Supplementary Material includes all reference values, excitation energies and statistical values for all tested cases and density functional approximations.

\section*{Data Availability Statement}
The data that supports the findings of this study are available within the article and its supplementary material.

\section*{Acknowledgements}
The authors would like to thank Dr Tobias Schwabe for the CCSD and CC3 values for the triplet excitations of the updated Gordon benchmark set which was obtained as part of our previous work published in 2017. M.C.-P. acknowledges a ‘Melbourne Research 
Scholarship’ by The University of Melbourne. L. G. acknowledges generous allocation of resources by the National Computational Infrastructure (NCI) National Facility within the National Computational Merit Allocation Scheme (project ID: fk5) and Research 
Platform Services (ResPlat) at The University of Melbourne (Project No. punim0094). This research was also supported by the sustaining and strengthening merit-based access to the NCI LIEF Grant (LE190100021) facilitated by The University of Melbourne.

\bibliographystyle{jcp}
\bibliography{LCTDDHDFs_triplets_paper}

\end{document}